\documentstyle[12pt,epsf]{article}

\newtheorem{theorem}{Theorem}[section]

\newtheorem{lemma}{Lemma}[section]

\newcommand{\mathcom}[3]
{\newcommand{#1}[#2]{\ifmmode#3\else \ifhmode $#3$\else #3\fi\fi}}

\mathcom{\myprime}{0}{^{\prime}}
\mathcom{\shiftprime}{0}{\,^{\prime}}

\begin{document}                                 
\nocite{*}

\title{\begin{flushright} 
          \begin{small} PUPT-1588 \end{small} 
       \end{flushright}
       A Geometric Formulation of Occam's Razor For Inference of
       Parametric Distributions} 
\author{Vijay Balasubramanian\thanks{vijayb@phoenix.princeton.edu} \\ 
        {\it Dept. of Physics, Princeton University, Princeton, NJ 08544}}

\date{\today}
\maketitle
                                 
                                 
\begin{abstract}                                 
I define a natural measure of the complexity of a parametric
distribution relative to a given true distribution called the {\it
razor} of a model family. The Minimum Description Length principle
(MDL) and Bayesian inference are shown to give empirical
approximations of the razor via an analysis that significantly extends
existing results on the asymptotics of Bayesian model selection.  I
treat parametric families as manifolds embedded in the space of
distributions and derive a canonical metric and a measure on the
parameter manifold by appealing to the classical theory of hypothesis
testing.  I find that the Fisher information is the natural measure of
distance, and give a novel justification for a choice of Jeffreys
prior for Bayesian inference.  The results of this paper suggest
corrections to MDL that can be important for model selection with a
small amount of data.  These corrections are interpreted as natural
measures of the simplicity of a model family.  I show that in a
certain sense the logarithm of the Bayesian posterior converges to the
logarithm of the {\it razor} of a model family as defined here.  Close
connections with known results on density estimation and ``information
geometry'' are discussed as they arise.

\end{abstract}

\section{Introduction}                                
\label{sec:intro}                                
William of Ockham, a great lover of simple explanations, wrote that
``a plurality is never to be posited except where
necessary.''\cite{maurer} The aim of this paper is to provide a
geometric insight into this principle of economy of thought in the
context of inference of parametric distributions.  The task of
inferring parametric models is often divided into two parts.  First of
all, a parametric family must be chosen and then parameters must be
estimated from the available data.  Once a model family is specified,
the problem of parameter estimation, although hard, is well understood
- the typical difficulties involve the presence of misleading local
minima in the error surfaces associated with different inference
procedures.  However, less is known about the task of picking a model
family, and practitioners generally employ a judicious combination of
folklore, intuition, and prior knowledge to arrive at suitable models.
                                 
The most important principled techniques that are used for model
selection are Bayesian inference and the Minimum Description Length
principle.  In this paper I will provide a geometric insight into both
of these methods and I will show how they are related to each other.
In Section~\ref{sec:qual} I give a qualitative discussion of the
meaning of ``simplicity'' in the context of model inference and
discuss why schemes that favour simple models are desirable.  In
Section~\ref{sec:deriv} I will analyze the typical behaviour of Bayes
rule to construct a quantity that will turn out to be a {\it razor} or
an index of the simplicity and accuracy of a parametric distribution
as a model of a given true distribution.  In effect, the razor will be
shown to be to be an ideal measure of ``distance" between a model
family and a true distribution in the context of parsimonious model
selection.  In order to define this index it is necessary to have a
notion of measure and of metric on a parameter manifold viewed as a
subspace of the space of probability distributions.
Section~\ref{sec:geom} is devoted to a derivation of a canonical
metric and measure on a parameter manifold.  I show that the natural
distance on a parameter manifold in the context of model inference is
the Fisher Information.  The resulting integration measure on the
parameters is equivalent to a choice of Jeffreys prior in a Bayesian
interpretation of model selection.  The derivation of Jeffreys prior
in this paper makes no reference to the Minimum Description Length
principle or to coding arguments and arises entirely from geometric
considerations. In a certain novel sense Jeffreys prior is seen to be
the prior on a parameter manifold that is induced by a uniform prior
on the space of distributions. Some relationships with the work of
Amari et.al. in information geometry are described.(\cite{amari85},
\cite{amari87}) In Section~\ref{sec:largen} the behaviour of the razor
is analyzed to show that empirical approximations to this quantity
will enable parsimonious inference schemes.  I show in
Section~\ref{sec:meaning} that Bayesian inference and the Minimum
Description Length principle are empirical approximations of the
razor.  The analysis of this section also reveals corrections to MDL
that become relevant when comparing models given a small amount of
data.  These corrections have the pleasing interpretation of being
measures of the robustness of the model.  Examination of the behaviour
of the razor also points the way towards certain geometric refinements
to the information asymptotics of Bayes Rule derived by Clarke and
Barron.(\cite{clarke}) Close connections with the index of
resolvability introduced by Barron and Cover are also
discussed.(\cite{barron})

\section{What is Simplicity?}                                
\label{sec:qual}                                 
                                 
\mathcom{\true}{0}{t}                                 
\mathcom{\events}{0}{E}                                 
\mathcom{\model}{0}{M}                                 
\mathcom{\modela}{0}{\model_1}                                 
\mathcom{\modelb}{0}{\model_2}                                 
\mathcom{\structure}{0}{S}                                 
\mathcom{\parameters}{0}{\Theta}                                 

\begin{figure}                                 
\begin{center}                                 
\leavevmode                                 
\epsfxsize=4in                                 
\epsffile{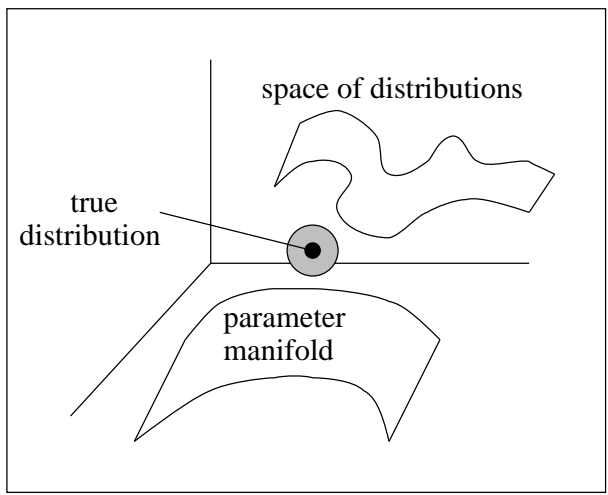}                                 
\end{center}                                 
\caption{Parameter Manifolds in The Space of Distributions\label{fig1}}                                 
\end{figure}

Since the goal of this paper is to derive a geometric notion of
simplicity of a model family it is useful to begin by asking why we
would wish to bias our inference procedures towards simple models.  We
should also ask what the qualitative meaning of ``simplicity'' should
be in the context of inference of parametric distributions so that we
can see whether the precise results arrived at later are in accord
with our intuitions.  For concreteness let us suppose that we are
given a set of $N$ outcomes $\events = \{e_1 \cdots e_N \} $ generated
i.i.d. from a true distribution~\true.  In some suitable sense, the
empirical distribution of these events will fall with high probability
within some ball around~\true~in the space of distributions. (See
Figure~\ref{fig1}.)  Now let us suppose that we are trying to
model~\true~with one of two parametric families~\modela~or~\modelb.
Now~\modela~and~\modelb~ define manifolds embedded in the space of
distributions (see Figure~\ref{fig1}) and the inference task is to
pick the distribution on~\modela~or~\modelb~that best describes the
true distribution.
                 
\begin{figure}                                 
\begin{center}                                 
\leavevmode                                 
\epsfysize=3in                                 
\epsffile{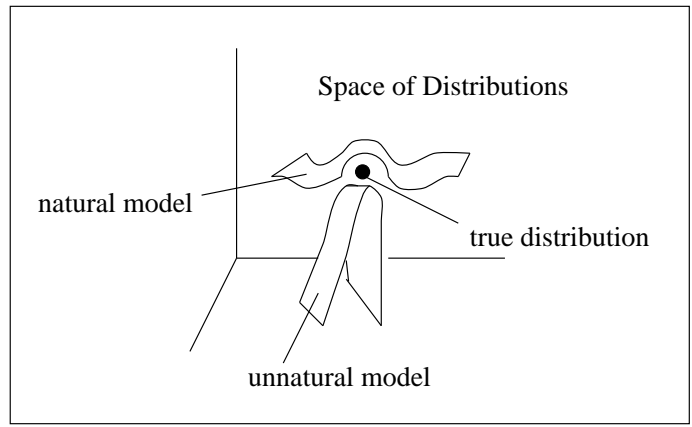}                                 
\end{center}                                 
\caption{Natural and Unnatural Models\label{fig2}}                                 
\end{figure}

   If we had an infinite number of outcomes and an arbitrary amount of
time with which to perform the inference, the question of simplicity
would not arise.  Indeed, we would simply use a consistent parameter
estimation procedure to pick the model distribution on~\modela~
or~\modelb~that gives the best description of the empirical data and
that would be guaranteed to give the best model of the true
distribution.  However, since we only have finite computational
resources and since the empirical distribution for finite $N$ only
approximates the true, our inference procedure has to be more careful.
Indeed, we are naturally led to prefer models with fewer degrees of
freedom.  First of all, smaller models will require less computational
time to manipulate.  They will also be easier to optimize since they
will generically have fewer misleading local minima in the error
surfaces associated with the estimation.  Finally, a model with fewer
degrees of freedom generically will be less able to fit statistical
artifacts in small data sets and will therefore be less prone to
so-called ``generalization error''.  Another, more subtle, preference
regarding models inferred from finite data sets has to do with the
``naturalness'' of the model.  Suppose we are using a family~\model~to
describe a set of $N$ outcomes drawn from~\true.  If the accuracy of
the description depends very sensitively on the precise choice of
parameters then it is likely that the true distribution will be poorly
modelled by~\model.(See Figure~\ref{fig2}.)  This is for two reasons -
1) the optimal choice of parameters will be hard to find if the model
is too sensitive to the choice, and 2) even if we succeed in getting a
good description of one set of sample outcomes, the parameter sensitivity
suggests that another sample will be poorly described.  In geometric
terms, we would prefer model families which describe a set of
distributions all of which are close to the true. (See
Figure~\ref{fig2}.) In a sense this property would make a family a
more ``natural'' model of the true distribution~\true~than another
which approaches~\true~very closely at an isolated point.
                                 
The discussion above suggests that for practical reasons inference
schemes operating with a finite number of sample outcomes should
prefer models that give good descriptions of the empirical data, have
fewer degrees of freedom and are ``natural'' in the sense discussed
above.  I will refer to the first property (good description) as {\it
accuracy} and the latter two (fewer degrees of freedom and
naturalness) as {\it simplicity}.  We will see that both accuracy and
simplicity of parametric models can be understood in terms of the
geometry of the model manifold in the space of distributions.  This
geometric understanding provides an interesting complement to the
minimum description length approach, which gives an implicit
definition of simplicity in terms of shortest description length of
the data and model.

\section{Construction of The Razor Of A Model}                                
\label{sec:deriv}                
The previous section has discussed  the qualitative meaning                                
of simplicity and its practical importance for inference of                                
distributions from a finite amount of data.    In this section we will                                 
construct a quantity that is an index of the accuracy and the simplicity                                 
of a model family as a description of a given true distribution.   We                                 
will show in later sections that empirical approximations of this                                 
quantity which we call the {\it razor} of a model will enable                                 
consistent and parsimonious inference of parametric probability distributions.                                 
                                 
\subsection{Construction From Bayes Rule}                                 
We will now motivate the definition of the razor via a construction
from the Bayesian approach to model inference.  (In later sections we
will conduct a more precise analysis of the relationship between the
razor and Bayes Rule.)  Suppose we are given a collection of outcomes
$\events = \{ e_1 \ldots e_N \}, \, e_i \in X $ drawn independently
from a true density~\true, defined with respect to Lebesgue measure on
$X$.  Suppose also that we are given two parametric families of
distributions A and B and we wish to pick one of them as the model
family that we will use.  The Bayesian approach to this problem
consists of computing the posterior conditional probabilities $\Pr(A |
\events)$ and $\Pr(B | \events)$ and picking the family with the
higher probability.  The conditional probabilities depend, of course,
on the specific outcomes, and so in order to understand the most
likely result of an application of Bayes Rule we should analyze the
statistics of the posterior probabilities.  Let A be parametrized by a
set of parameters $\Theta = \{\theta_1, \ldots \theta_d \}$.  Then
Bayes Rule tells us that:
\begin{equation}                                 
\label{eq:bayes1}                                 
\Pr(A | \events) = \frac{\Pr(A)}{\Pr(\events)} \int d\mu(\Theta) \, \,
             w(\Theta) \Pr(\events |\Theta)
\end{equation}                                 
In this expression $\Pr(A)$ is the prior probability of the model
family, $w(\Theta)$ is a prior density with respect to Lebesgue
measure on the parameter space and $Pr(\events)$ is a prior density on
the $N$ outcome sample space.  The Lebesgue measure induced by the
parametrization of the $d$ dimensional parameter manifold is denoted
$d\mu(\Theta)$.  Since we are interested in comparing $\Pr(A |
\events)$ with $\Pr(B | \events)$, the prior $\Pr(\events)$ is a
common factor that we may omit and for lack of any better choice we
take the prior probabilities of A and B to be equal and omit them.  In
order to analyze the typical behaviour of Equation~\ref{eq:bayes1}
observe that $\Pr(\events | \Theta) = \prod_{i=1}^{N} \Pr(e_i |
\Theta) = \exp \left[ \sum_{i=1}^{N} \ln\Pr(e_i|\Theta) \right] $.
Define $G_i(\Theta) = \ln\Pr(e_i|\Theta)$ and $F(\Theta) =
\sum_{i=1}^N G_i(\Theta)$.  We see that $F(\Theta)$ is the sum of $N$
identically distributed, independent random variables.  Consequently,
as $N$ grows large the Central Limit Theorem applies and we can write
down the probability distribution for $F$ as:
\begin{equation}                                 
\Pr\left(F(\Theta) = f\right) \rightarrow \frac{1}{\sqrt{2 \pi N
    \sigma^2}} \exp-\frac{(f - N \mu)^2}{2 N \sigma^2}
\end{equation}                                 
where $\mu$ and $\sigma$ are the mean and standard deviation of the
$G_i$ in the true distribution and are defined as $\mu =
<G_i(\Theta)>_t = \int dx \, t(x) \ln\Pr(x|\Theta)$ and $\sigma^2 =
<G_i(\Theta)^2>_t - <G_i(\Theta)>_t^2$.  The most likely value of
$F(\Theta)$ is $N \mu$ and $\mu$ can be written in the following
pleasing form:
\begin{equation}                                 
<G_i(\Theta)>_t = -\int dx \, t(x)
                 \ln\left(\frac{t(x)}{\Pr(x|\Theta)}\right) + \int dx
                 \, t(x) \ln\left(t(x)\right) = -D(t \| \Theta) - h(t)
\end{equation}                                 
where $h(t)$, the differential entropy of the true distribution, is
assumed finite, and $D(t \| \Theta)$ is the relative entropy or
Kullback-Liebler distance between $t$ and the distribution indexed by
$\Theta$.  This suggests that the following quantity is worthy of
investigation:
\begin{equation}                                 
\label{eq:cand}                    
R_N^C(A) \propto \int d\mu(\Theta) \, w(\Theta) \exp{-N\left(
D(t\|\Theta) - h(t) \right)}
\end{equation}                 
(The superscript $C$ is intended to indicate that
Equation~\ref{eq:cand} is a {\it candidate} razor that we will improve
in the subsequent discussion.)  We will see in
Section~\ref{sec:relbayes} that $R_N^C(A)$ is closely related to the
typical asymptotics of $\ln\Pr(E|A)$.  In the business of comparing
$\Pr(A|\events)$ and $\Pr(B|\events)$, $\exp{-N h(t)}$ is a common
factor.  So we drop it and also note that in the absence of any prior
information the most conservative choice for $w(\Theta)$ appears to be
the uniform prior on the parameter manifold. (We will return to
examine this point critically and we will find that the natural prior
is not in fact uniform in the parameters.)  So we finally write our
candidate razor as:
\begin{equation}                                 
\label{eq:raz1}                                 
R_N^C(A) = \Pr(A | \events) = \frac{\int d\mu(\Theta) \exp{-N
                        D(t\|\Theta)}} {\int d\mu(\Theta)}
\end{equation}                                 
We have assumed a compact parameter manifold so that the uniform
distribution on the surface can be written as one over the volume.  We
take the integration measure $d\mu(\Theta)$ to be the Lebesgue measure
induced on the manifold by the atlas defined via the parametrization.
These definitions can be extended to non-compact parameter manifolds
with a little bit of care, but we will not do this here. The quantity
$R_N^C(A)$ defined in Equation~\ref{eq:raz1} is our candidate for a
natural meaure of the accuracy as well as the simplicity of a
parametric model distribution.  The construction of the razor in this
section is intended to be motivational.  We will see in
Section~\ref{sec:relbayes} that the razor is closely related to the
typical asymptotics of the logarithm of $\Pr(A|E)$.  Note that the
razor is not a quantity that is estimated from data - it is a
theoretical measure of complexity like the ``index of resolvability''
introduced by Barron and Cover and discussed in
Section~\ref{sec:mincomp}.(\cite{barron}) We will show that an
accurate estimator of the razor can be used to implement consistent
and parsimonious inference of probability distributions.
                                 
\subsection{A Difficulty}                                 
There is a major difficulty with an interpretation of $R_N^C(A)$ in
Equation~\ref{eq:raz1} as an intrinsic measure of qualities such as
the simplicity of a parametric family.  This difficulty arises because
we have not defined the integration measure sufficiently carefully. To
see this in pedestrian terms, consider a family with two parameters,
$x$ and $y$, for which the naive integration measure in the razor
would be $d\mu(\Theta) = dx \, dy$.  We could do the integration in
polar coordinates ($r$ and $\phi$), in which case the measure would be
$d\mu(\Theta) = r \, dr \, d\phi$.  Now the model could have been
specified in the first place in terms of the coordinates $r$ and
$\phi$ in which case the naive integration measure would have been
$d\mu(\Theta) = dr \, d\phi$ which will clearly yield a different
definition of the razor.  In other words, the razor as defined above
is not reparametrization invariant and consequently measures something
about both the model family and its parametrization.  In order to
define the razor as an intrinsic measure of the simplicity of a
parametric distribution we need to have an invariant integration
measure on the parameter manifold.  This is easily achieved - if we
know how to introduce a metric on the surface, the metric will induce
a measure with required properties.  But what is the correct metric on
the parameter manifold?  Since the model is embedded in the space of
distributions, the metric on the manifold should be induced from a
natural distance in the space of distributions.
                                 
   Further insight into this issue is obtained by considering the
Bayesian construction of the razor.  In the course of this
construction we assumed a uniform prior on the parameter manifold.
Now the manifold itself is some parameter invariant object that lives
in the space of distributions. Let A be a set of distributions in the
parameter manifold.  The uniform prior associated with different
parametrizations will assign different measures to the set A.  On the
one hand we could say that the choice of parametrization of a model
involves an implicit choice of measure on the manifold and that a
parameter dependence is therefore to be expected in Bayesian methods.
However, it seems more correct to say that we did not actually mean to
say that all {\it parameters} are equally likely - our intention was
to say that in the absence of any prior information, all {\it
distributions} are equally likely.  In other words, to apply Bayesian
methods properly to the task of model inference, we have to find a way
of assigning a uniform prior in the space of distributions and induce
from that a measure on parameter manifolds.

In the next section we will use the observations made in the previous
paragraphs to derive a metric and a measure on the parameter manifold
that make the razor a parameter-invariant measure of the simplicity of
a model.  We will find that the natural metric on the parameter
manifold is the Fisher Information on the surface and the
reparametrization invariant razor is consequently given by:
\begin{equation}                                 
\label{eq:raz2}                                 
R_N(A) = \frac{\int d\mu(\Theta) \sqrt{\det{J}} \exp{-N D(t||\Theta)}}                                 
              {\int d\mu(\Theta) \sqrt{\det{J}}}                                 
\end{equation}                                 
where $J$ is the Fisher Information matrix.  The work of Rao, Fisher,
Amari and others has previously suggested this choice of measure and
metric.(\cite{amari85}, \cite{amari87}).  However, as pointed out by
these authors, there are many potential choices of metrics on
parameter manifolds and the choice of a metric requires careful
justification.  Once a metric is chosen the standard apparatus of
differential geometry may be unfolded and statistical interpretations
can be attached to geometric quantitites.  In the following sections I
provide justifications for the choice of the Fisher Information as the
metric appropriate to model estimation.
                                 
The choice of $\sqrt{\det{J}}$ as the integration measure is
equivalent to a choice of Jeffreys prior in the Bayesian
interpretation of the razor.(\cite{lee}, \cite{jeffreys}) Jeffreys
recommends this choice of prior because, as we will note in later
sections, its definition guarantees the reparametrization invariance
of the Bayesian posterior.  However, the requirement of
reparametrization invariance alone does not uniquely fix the prior -
{\it any} prior which is defined to have suitable transformation
properties under reparametrizations of a model will yield the desired
invariance.\footnote{It is clear that any prior that is defined as the
square root of the determinant of a two form on the parameter manifold
will be invariant under reparametrizations.  So reparametrization
invariance is hardly sufficient to pick out a unique prior.} Indeed,
Jeffreys considers priors related to different distances on the space
of distributions including the $L^k$ norms and the relative entropy
distance.  I will show that choosing a Jeffreys prior is equivalent to
assuming equal prior likelihood of all {\it distributions} as opposed
to equal prior likelihood of parameters.
                                 
 From the point of view of statistical mechanics the razor can be
interpreted as a partition function with energies given by the
relative entropy and temperature $1/N$.  Since temperature regulates
the size of fluctuations in physical systems as does the number of
events in statistical systems, this analogy makes good sense.  In
Section~\ref{sec:largen} we exploit the techniques of statistical
mechanics to develop a systematic series expansion for the razor. The
geometrical interpretation will become more clear as the reader
proceeds further.

\section{Geometry of Parameter Manifolds}                                
\label{sec:geom}                                 
In this section I will derive a natural metric and measure on a                                 
parameter manifold.  We will see that the Fisher Information is the                                 
natural metric and the natural measure is associated with this metric.                                 
The Fisher Information has a long history as a local measure of                                 
distance in the space of distributions starting with the Cramer-Rao                                 
bounds.  The work of Fisher, Rao, Amari and others has elucidated the                                 
role of geometry in statistics (\cite{amari85},\cite{amari87}) and                                 
there is a sizable literature on the construction and interpretation                                 
of geometric quantities in information theory.  However, since there                                 
are many potential metrics in the space of distributions the important                                 
issue is to determine which metric is appropriate to a given problem.                                 
Once this is done, the theory of Riemannian manifolds provides the                                 
necessary technology for manipulating parametric families in the space                                 
of distributions and the difficult task is to identify the geometric                                 
quantities of interest to statistics. In this section we will present                                 
two derivations of the natural integration measure on a parameter                                 
manifold in the context of density estimation.                                 
                                 
\subsection{Distance Induced By The Relative Entropy}                                 
\label{sec:reldist}                                 
It is useful to start with a somewhat heuristic derivation that                                 
recapitulates arguments made in the ``information geometry''                                 
literature. (\cite{amari85}, \cite{amari87})  Let us                                  
start by assuming that in the context of model inference, the natural                                 
distance on the space of distributions is the relative entropy $D(p\|q)$.                                 
Unfortunately, $D$ does not define a metric since it is not symmetric                                 
and does not obey triangle inequalities except in special cases.                                 
However, as we shall see, $D(p\|q)$ will induce a Riemannian metric on a                                 
parameter manifold given suitable technical conditions.  Let $\cal{M}$                                 
be a manifold in the space of distributions with local coordinates                                 
$\Theta = \{ \theta_1 \ldots \theta_d \} $.  Let $p$ be a fixed point                                 
on $\cal{M}$ and $q$ be any other point.   Then the relative entropy                                 
between $p$ and $q$, $D(\Theta_p \| \Theta_q)$ is a non-negative                                 
function of $\Theta_q$ that attains its minimum at $\Theta_q = \Theta_p$                  
and the value of the minimum is zero.  This means that the zeroth and first order terms in                 
the Taylor  expansion of $D(\Theta_p \| \Theta_q)$ at $p$ vanish identically.                                 
Letting $\Delta\Theta = \Theta_q - \Theta_p$, and assuming twice                                 
differentiability of the relative entropy in a neighbourhood of $p$,                                 
we can Taylor expand to second order:                                 
\begin{eqnarray}                                 
\label{eq:taylor}                                 
D(\Theta_p \|  \Theta_q) \approx                                  
- \int dx \: \Pr(x|\Theta_p) & \frac{1}{2}\left[ \frac{1}{\Pr(x|\Theta_p)}                                 
                             \frac{\partial}{\partial\theta_i}                                 
                             \frac{\partial}{\partial\theta_j}                                 
                             \Pr(x|\Theta_p) \right.  -                                 
\nonumber  \\                                  
                     &      \left.   \frac{1}{{\Pr(x|\Theta_p)}^2}                                 
                       \frac{\partial\Pr(x|\Theta_p)}{\partial\theta_i}                                 
                       \frac{\partial\Pr(x|\Theta_q)}{\partial\theta_j}                                 
                              \right] \Delta\Theta^i \Delta\Theta^j                                 
\end{eqnarray}                                 
In the above equation as in all future equations, repeated indices are                                 
implicitly summed over.  So, for example, there is an implicit sum on                                 
$i$ and $j$ in Equation~\ref{eq:taylor}.   The first term in this                                 
equation vanishes if the derivatives with respect to $\theta_i$ and                                 
$\theta_j$  commute with the integral\footnote{$\int dx \partial_{\theta_i}                                 
\partial_{\theta_j} \Pr(x|\Theta) = \partial_{\theta_i}                                 
\partial_{\theta_j} \int dx \Pr(x|\Theta) = \partial_{\theta_i}                                 
\partial_{\theta_j}  1 = 0$}.    We assume this commutativity and                                 
recognize the remaining term as one-half times the Fisher Information on the parameter                                 
manifold $D(\Theta_p \| \Theta_q) = (1/2) < \partial_{\theta_i}                                 
\ln{\Pr(x|\theta_p)} \partial_{\theta_j} \ln{\Pr(x|\theta_p)}                                 
>_{\Theta_p}  \Delta\theta^i \Delta\theta^j = (1/2) J_{ij} \Delta\theta^i                                 
\Delta\theta^j$.\footnote{The same result can be arrived at by looking                                 
at the second order Taylor expansion around $p$ of the  symmetrized                                 
relative entropy   $D(p \| q) + D(q \| p)$.   In this case there is no                                 
need to make any further assumptions about commutativity of the                                 
derivatives and integral.}                                   
                                 
  We have found that if we accept that the relative entropy is the
natural measure of distance between distributions in the context of
model estimation, the induced distance between nearby points on a
parameter manifold is $D(p,q) = (1/2) J_{ij} \Delta\theta^i
\Delta\theta^j$ to leading order in $\Delta\theta$.  Since the Fisher
Information appears in this expression as a quadratic form, it is
tempting to interpret it as the natural metric on the surface.  We
will only consider consider models where the determinant of the Fisher
Information is non-vanishing everwhere on the surface.  This
non-degeneracy condition essentially guarantees that nearby points on
a model manifold describe sufficiently different distributions.  Since
we derived the Fisher Information metric from a Taylor expansion at
the minimum of a function we conclude that for the non-degenerate
models that are of interest to us, the Fisher Information is a
positive definite metric on the model manifold.  Therefore, we can
appeal to the standard theory of Riemannian geometry to observe that
the reparametrization invariant integration measure on the manifold is
$\sqrt{\det{J}}$ where $J$ is the Fisher Information.  Putting this
measure into Equation~\ref{eq:raz1} for the razor, we immediately get
Equation~\ref{eq:raz2} which is now coordinate-independent and a
candidate for a measure of some intrinsic properties of a parametric
distribution.
                                 
\subsection{How To Count Distinguishable Models}                                 
\label{sec:distinguishable}                                 
The Bayesian derivation of the razor of a model provides good                                 
intuitions for a more careful derivation of the integration measure.                                 
As we have discussed, in the Bayesian                                 
interpretation we would like to say that all distributions are equally                                 
likely.  If this is the case we should give equal weight to all                                 
distinguishable distributions on a model manifold.  However, nearby                                 
parameters index very similar distributions.  So let us ask the                                 
question, ``How do we count the number of distinct distributions in                                 
the neighbourhood of a point on a parameter manifold?''  Essentially,                                 
this is a question about the embedding of the parameter manifold in                                 
the space of distributions.  Points that are distinguishable as                                 
elements of $R^n$ may be mapped to indistinguishable points (in some                                 
suitable sense) of the embedding space.

To answer the question let us take $p$ and $q$ to be points on a
parameter manifold.  Since we are working in the context of density
estimation a suitable measure of the distinguishability of $\Theta_p$
and $\Theta_q$ should be derived by taking $N$ data points drawn from
either $p$ or $q$ and asking how well we can guess which distribution
produced the data.  If $p$ and $q$ do not give very distinguishable
distributions, they should not be counted separately in the razor
since that would count the same distribution twice.
                                 
   Precisely this question of distinguishability is addressed in the
classical theory of hypothesis testing.  Suppose $\{e_1 \ldots e_N
\} \in E^N$ are drawn iid from one of $f_1$ and $f_2$ with
$D(f_1\|f_2) < \infty$. Let $A_N \subseteq E^N$ be the acceptance
region for the hypothesis that the distribution is $f_1$ and define
the error probabilities $\alpha_N = f_1^N(A_N^C)$ and $\beta_N =
f_2^N(A_N)$.  ($A_N^C$ is the complement of $A_N$ in $E^N$ and $f^N$
denotes the product distribution on $E^N$ describing $N$ iid outcomes
drawn from $f$.)  In these definitions $\alpha_N$ is the probability
that $f_1$ was mistaken for $f_2$ and $\beta_N$ is the probability of
the opposite error.  Stein's Lemma tells us how low we can make
$\beta_N$ given a particular value of $\alpha_N$.  Indeed, let us
define:
\begin{equation}                                 
\beta_N^\epsilon = \min_{\stackrel{A_N \subseteq E^N}{\alpha_N \leq                                 
\epsilon}}  \beta_N                                                     
\end{equation}                                 
Then Stein's Lemma tells us that:                                 
\begin{equation}                                 
\lim_{\epsilon \rightarrow 0} \lim_{N \rightarrow \infty} \frac{1}{N}                                 
\ln \beta_N^\epsilon = - D(f_1 \| f_2)                                  
\end{equation}                                 
By examining the proof of Stein's Lemma (\cite{cover}) we find that
for fixed $\epsilon$ and sufficiently large $N$ the optimal choice of
decision region places the following bound on $\beta_N^\epsilon$:
\begin{equation}                                 
-D(f_1\|f_2) - \delta_N + \frac{\ln{(1 - \alpha_N)}}{N} \leq                                 
\frac{1}{N} \ln\beta_N^\epsilon \leq                                 
-D(f_1\|f_2) + \delta_N + \frac{\ln{(1 - \alpha_N)}}{N}                                  
\end{equation}                                 
where $\alpha_N < \epsilon$ for sufficiently large $N$.  The
$\delta_N$ are any sequence of positive constants that satisfy the
property that:
\begin{equation}         
\label{eq:deltas}         
\alpha_N = f_1^N(|\frac{1}{N}\sum_{i=1}^N
\ln{\frac{f_1(e_i)}{f_2(e_i)}} - D(f_1\|f_2) | > \delta_N) \leq
\epsilon
\end{equation}         
for all sufficiently large $N$.  The strong law of large number
numbers tells us that $(1/N) \sum_{i=1}^N \ln(f_1(e_i)/f_2(e_i))$
converges to $D(f_1\|f_2)$ almost surely since $D(f_1\|f_2) =
E_{f_1}(\ln(f_1(e_i)/f_2(e_i))$.  Almost sure convergence implies
convergence in probability so that for any fixed $\delta$ we have:
\begin{equation}         
\label{eq:deltas2}         
f_1^N(|\frac{1}{N}\sum_{i=1}^N \ln{\frac{f_1(e_i)}{f_2(e_i)}} -
D(f_1\|f_2) | > \delta) < \epsilon
\end{equation}         
for all sufficiently large N.  For a fixed $\epsilon$ and a fixed $N$
let $\Delta_{\epsilon,N}$ be the collection of $\delta > 0$ which
satisfy Equation~\ref{eq:deltas2}.  Let $\delta_{\epsilon N}$ be the
infimum of the set $\Delta_{\epsilon,N}$.  Equation~\ref{eq:deltas2}
guarantees that for any $\delta >0$, for any sufficiently large $N$,
$0 < \delta_N < \delta$.  We conclude that $\delta_{\epsilon N}$
chosen in this way is a sequence that converges to zero as $N
\rightarrow \infty$ while satisfying the condition in
Equation~\ref{eq:deltas} which is necessary for proving Stein's Lemma.
We will now apply these facts to the problem of distinguishability of
points on a parameter manifold.
                                 
Let $\Theta_p$ and $\Theta_q$ index two distributions on a parameter manifold and        
suppose that we are given $N$ outcomes generated independently from one of them.       
We are interested in using Stein's Lemma to determine how distinguishable $\Theta_p$        
and $\Theta_q$ are.   By Stein's Lemma:       
\begin{equation}       
\label{eq:st1}       
-D(\Theta_p\|\Theta_q) - \delta_{\epsilon N}(\Theta_q) + \frac{\ln(1 - \alpha_N)}{N}       
\leq       
\frac{\beta_N^\epsilon(\Theta_q)}{N}       
\leq       
-D(\Theta_p\|\Theta_q) + \delta_{\epsilon N}(\Theta_q) + \frac{\ln(1 - \alpha_N)}{N}       
\end{equation}       
where we have written $\delta_{\epsilon N}(\Theta_q)$ and
$\beta_N^\epsilon(\Theta_q)$ to emphasize that these quantities are
functions of $\Theta_q$ for a fixed $\Theta_p$.  Let $A =
-D(\Theta_p\|\Theta_q) + (1/N)\ln(1- \alpha_N)$ be the average of the
upper and lower bounds in Equation~\ref{eq:st1}.  Then $A \geq
-D(\Theta_p\|\Theta_q) + (1/N) \ln(1 - \epsilon)$ because the
$\delta_{\epsilon N}(\Theta_q)$ have been chosen to satisfy
Equation~\ref{eq:deltas}.  We now define the set of distributions $U_N
= \{\Theta_q : -D(\Theta_p\|\Theta_q) + (1/N) \ln(1 - \epsilon) \geq
(1/N) \ln{\beta^*} \}$ where $1 > \beta^* > 0$ is some fixed constant.
Note that as $N \rightarrow \infty$, $D(\Theta_p\|\Theta_q)
\rightarrow 0$ for $\Theta_q \in U_N$.  We want to show that $U_N$ is
a set of distributions which cannot be very well distinguished from
$\Theta_p$.  The first way to see this is to observe that the average
of the upper and lower bounds on $\ln{\beta^\epsilon_N}$ is greater
than or equal to $\ln\beta^*$ for $\Theta_q \in U_N$.  So, in this
loose, average sense, the error probability $\beta_N^\epsilon$ exceeds
$\beta^*$ for $\Theta_q \in U_N$.  More carefully, note that $(1/N)
\ln(1 - \alpha_N) \geq (1/N) \ln(1 - \epsilon)$ by choice of the
$\delta_{\epsilon N}(\Theta_q)$.  So, using Equation~\ref{eq:st1} we
see that $(1/N) \ln{\beta_N^\epsilon(\Theta_q)} \geq (1/N)
\ln{\beta^*} - \delta_{\epsilon N}(\Theta_q)$.  Exponentiating this
inequality we find that:
\begin{equation}       
\label{eq:expbound}       
1       
\geq        
\left[ \beta_N^\epsilon(\Theta_q) \right]^{(1/N)}        
\geq       
(\beta^*)^{(1/N)} \, e^{-\delta_{\epsilon N}(\Theta_q)}       
\end{equation}       
The significance of this expression is best understood by considering
parametric families in which, for every $\Theta_q$, $X_q(e_i) =
\ln(\Theta_p(e_i)/\Theta_q(e_i))$ is a random variable with finite
mean and bounded variance, in the distribution indexed by $\Theta_p$.
In that case, taking $b$ to be the bound on the variances,
Chebyshev's inequality says that:
\begin{equation}       
\Theta_p^N\left(|\frac{1}{N} \sum_{i=1}^N X_q(e_i) \, - \, D(\Theta_p\|\Theta_q) | > \delta        
\right)        
\leq        
\frac{Var(X)}{\delta^2 \, N}        
\leq        
\frac{b}{\delta^2 \, N}       
\end{equation}       
In order to satisy $\alpha_N \leq \epsilon$ it suffices to choose
$\delta = (b/N\epsilon)^{1/2}$.  So, if the bounded variance condition
is satisfied, $\delta_{\epsilon N}(\Theta_q) \leq (b/N\epsilon)^{1/2}$
for any $\Theta_q$ and therefore we have the limit $\lim_{N
\rightarrow \infty} \sup_{\Theta_q \in U_N} \delta_{\epsilon
N}(\Theta_q) = 0$.  Applying this limit to Equation~\ref{eq:expbound}
we find that:
\begin{equation}       
\label{eq:asstein}       
1        
\geq       
\lim_{N \rightarrow \infty} \inf_{\Theta_q \in U_N} \left[\beta_N^\epsilon(\Theta_q)        
\right]^{(1/N)} \geq 1 \times \lim_{N \rightarrow \infty} \inf_{\Theta_q \in U_N}       
e ^{-\delta_{\epsilon N}(\Theta_q)} = 1        
\end{equation}       
In summary we find that $\lim_{N \rightarrow \infty} \inf_{\Theta_q
\in U_N} [ \beta_N^\epsilon(\Theta_q) ]^{(1/N)} = 1$.  This is to be
contrasted with the behaviour of $\beta_N^\epsilon(\Theta_q)$ for any
{\it fixed} $\Theta_q \neq \Theta_p$ for which $\lim_{N \rightarrow
\infty} [\beta_N^\epsilon(\Theta_q)]^{(1/N)} = \exp{-
D(\Theta_p\|\Theta_q)} < 1 $.  We have essentially shown that the sets
$U_N$ contain distributions that are not very distinguishable from
$\Theta_p$.  The smallest one-sided error probability
$\beta_N^\epsilon$ for distinguishing between $\Theta_p$ and $\Theta_q
\in U_N$ remains essentially constant leading to the asymptotics in
Equation~\ref{eq:asstein}.
       
Define $\kappa \equiv -\ln\beta^* + \ln(1-\epsilon)$ so that we can
summarize the region $U_N$ of high probability of error $\beta^*$ at
fixed $\epsilon$ as $\kappa/N \geq D(\theta_p\|\theta_q)$.\footnote{We
will eventually take the limits $N \rightarrow \infty$, $\epsilon
\rightarrow 0$ and $\beta^* \rightarrow 1$ in that order.}  As N grows
large for fixed $\kappa$, the distributions $\Theta_p$ and $\Theta_q$
must be close in relative entropy sense and so we can write $\Theta_q
= \Theta_p + \Delta\Theta$ and Taylor expand the relative entropy on
the manifold near $p$.  By arguments identical to those made in
Section~\ref{sec:reldist} we conclude that $D(\Theta_p \| \Theta_q) =
(1/2) J_{ij}(\Theta_p) \Delta\theta^i \Delta\theta^j +
O(\Delta\Theta^3)$ where, as before, we have used the index summation
convention and defined the Fisher Information $J_{ij}$ from the matrix
of second derivatives of the relative entropy.\footnote{See
Section~\ref{sec:reldist} for the assumptions concerning
differentiability and commutation of derivatives and integrals.}  So,
the nearly indistinguishable region $U_N$ around $\Theta_p$ is
summarized by $J_{ij}(\Theta_p) \Delta\theta^i \Delta\theta^j \leq 2
\kappa / N + O(\Delta\Theta^3)$, which defines the interior of an
ellipsoid on the parameter manifold.  For large $N$, $\kappa /N$ is
small, and so, since the manifold is locally Euclidean, the volume of
this ellipsoid is given by: \mathcom{\vol}{0}{V_{\epsilon,\beta^*, N}}
\begin{equation}                                 
\label{eq:volume}                                 
\vol =                                 
\left( 2\pi \frac{\kappa}{N} \right)^{d/2} \frac{1}{\Gamma(d/2 +                                 
1)}   \frac{1}{\sqrt{det J_{ij}}}                                 
\end{equation}                                 
We refer to~\vol~as the volume of indistinguishability at levels
$\epsilon$, $\beta^*$ and $N$.  It measures the volume of parameter
space in which the distributions are indistinguishable from $\Theta_p$
with error probabilities $\alpha_N \leq \epsilon$ and
$(\beta_N^\epsilon)^{(1/N)} \geq (\beta^*)^{(1/N)}
\exp{-\delta_{\epsilon N}}$, given $N$ sample events.

If $\beta^*$ is very close to one, the distributions inside~\vol~are
not very distinguishable and should not be counted separately in the
razor.  (Equivalently, the Bayesian prior should not treat them as
separate distributions.)  We wish to construct a measure on the
parameter manifold that reflects this indistinguishability.  We will
also assume a principle of ``translation invariance" in the space of
distributions by supposing that volumes of indistinguishability at
given values of $N$, $\beta^*$ and $\epsilon$ should have the same
measure regardless of where in the space of distributions they are
centered.
In what follows we will define a sequence of measures that reflect
indistinguishability and translation invariance at each level
$\beta^*$, $\epsilon$ and $N$ in the space of distributions.  The
continuum measure on the manifold is obtained by considering the
limits of integrals defined with respect to this sequence of measures.
We begin with the Lebesgue measure induced on the model manifold by
the parameter embedding in $R^d$.  For convenience we will assume that
the model manifold can be covered by a single parameter patch so that
issues of consistent sewing of patches do not arise.  A real function
on the model manifold will be called a {\it step} map with respect to
a finite, Lebesgue measurable partition of the manifold if it is
constant on each set in the partition.  For any Lebesgue measurable
function $f$ there is a sequence of step maps that converges pointwise
to $f$ almost everywhere and in $L^1$.(\cite{lang}) We will assume
that the Fisher Information matrix $J$ is non-singular everywhere and
is component-wise Lebesgue measurable.  The determinant of $J$ will
consequently be everywhere finite and also Lebesgue measurable.

Let $f$ be a step map with respect to some partition $A = \{A_i\}$ of
the model manifold and let $J_{ij}$ be a Fisher Information matrix
which is non-singular everywhere and a step map with respect to a
partition $B =\{B_i\}$.  Let $K= \{A_i \bigcap B_j \, :\, A_i \in A \,
, \, B_j \in B \}$ be a partition of the manifold such that if $K_i
\in K$ then both $f$ and $J$ are constant on $K_i$.  Consider fixed
values of $\beta^*$, $\epsilon$ and $N$ in the above definition of the
volumes of indistinguishability.  At fixed $\beta^*$, $\epsilon$ and
$N$ we would like to define a measure $\nu_{\epsilon\beta^* N}$ by
covering the sets $K_i \in K$ economically with volumes of
indistinguishability and placing delta functions at the center of each
volume in the cover.  Such a definition would give each volume of
indistinguishability equal weight in an integral over the model
manifold and would ignore variations in an integrand on a scale
smaller than these volumes.  As such, the definition would reflect the
properties of indistinguishability and translation invariance at fixed
$\beta^*$, $\epsilon$ and $N$.  As the volumes of indistinguishability
shrink we could hope to define a continuum limit of this discrete
sequence of measures.  The following discussion gives a careful
prescription for carrying out this agenda.  The argument should be
regarded as a ``construction'' consistent with the principles of
indistinguishability and translation invariance rather than as a
``derivation''.
    
    Since the program outlined above involves covering arbitrary
measurable subsets of $R^d$ with volumes of indistinguishability, we
begin by amassing some useful facts about covers of $R^d$ by spheres.
(See~\cite{conway}) Let $C_r$ be a cover of $R^d$ by spheres of radius
$r$.  Let $H \subset R^d$ have finite Lebesgue measure and let
$N_H(C_r)$ be the number of sphere in $C_r$ that intersect $H$.
Define the covering density of $H$ induced by $C_r$, $D(H,C_r)$ , to
be:
\begin{equation}  
\label{eq:dendef}  
D(H,C_r) = \frac{N_H(C_r) \, v_d(r)}{\mu(H)}  
\end{equation}  
where $v_d(r)$ is the volume of a $d$ dimensional sphere of radius $r$
and $\mu(H)$ is the Lebesgue measure of $H$.  Let $S_L$ be a square of
side $L$ centered at any point in $R^d$.  For a fixed covering radius
$r$, let $\tau = r/L$ and let $N_{S_L}$ be the number of spheres of
$C_r$ that intersect $S_L$.  Then define the covering density of $R^d$
induced by $C_r$ to be:
\begin{equation}  
\label{eq:rdens}  
D(R^d,C_r) = \lim_{\tau \rightarrow 0} \: D(S_L,C_r)   
=  
\lim_{\tau \rightarrow 0} \frac{v_d(r) N_{S_L}}{L^d}  
\end{equation}  
so long as this limit exists.  (Usually the limit $L\rightarrow
\infty$ is taken, but we will find the current formulation easier to
work with.)  Let a {\it minimal} cover of $R^d$ with spheres of radius
$r$ be a cover that attains the minimum possible $D(R^d,C_r)$ over all
covers $C_r$.  This minimal density is independent of $r$ and so we
will write it as $D(d)$.  To show this independence, suppose that
there is an $r$ dependence and that the minimal densities for $r_1$
and $r_2$ have the relationship $D(d,r_1) < D(d,r_2)$.  Then by
rescaling the coordinates of $R^d$ by $r_2/r_1$ we can convert the
cover by spheres of radius $r_1$ into a cover by spheres of radius
$r_2$.  However, Equation~\ref{eq:rdens} shows that the density of the
cover would remain unchanged since the sides of the squares $L$ would
increase in length by $r_2/r_1$.  This would give a cover with radius
$r_2$ whose density is less than the density $D(d,r_2)$ implying that
the latter density cannot be minimal.  It is well known that the
minimal density for covering $R^d$ with spheres, $D(d)$, is greater
than $1$ so that the volumes of indistinguishability used in
covering parameter manifolds will necessarily intersect each other.
We will pick the minimal cover using volumes of indistinguishability
in order to minimize overcounting of distributions in the measure that
will be derived via the construction presented in this paper.
  
The construction of a measure on a parameter manifold that respects
indistinguishability and translation invariance requires the property
that the density of the covering by $C_r$ of any disjoint union of
sufficiently large squares approaches $D(d)$ when the limit in
Equation~\ref{eq:rdens} exists.  Indeed, the following lemma is easy
to show:
\begin{lemma}  
\label{lemma:constr}  
Let $C_r$ be a covering of $R^d$ by spheres of radius $r$ for which
$D(R^d,C_r)$ exists.  Then for any $\epsilon > 0$ there is a $\tau_0 >
0$ such that if $S$ is a finite union of squares intersecting at most
on their boundaries and each of whose sides exceeds $L_0$ satisfying
$r/L_0 < \tau _0$, then $|D(S,C_r) - D(R^d,C_r)| < \epsilon$.
\end{lemma}  
{\bf Proof:} Let $S_L$ be a square of side $L$ and let $N_L$ be the
number of spheres in $C_r$ that intersect $S_L$.  Let $B_L$ be the
number of spheres that intersect the boundary of $S_L$.  Take
$S_{L-2r}$ to be a square of side $L-2r$, centered at the same
location as $S_L$.  Then $B_L \leq N_{L} - N_{L-2r}$.  By
Equation~\ref{eq:rdens}, for any $\epsilon^\prime$, we can pick $r/L$
to be small enough so that:
\begin{eqnarray}  
D - \epsilon^\prime &\leq \frac{N_{L-2r} \, v_d(r)}{(L-2r)^d} & \leq D
+ \epsilon^\prime \label{eq:asdf} \\ D - \epsilon^\prime &\leq
\frac{N_{L} \, v_d(r)}{(L)^d} & \leq D + \epsilon^\prime
\label{eq:asdf1}  
\end{eqnarray}  
where $D \equiv D(R^d,C_r)$. Writing $N_{L-2r} \leq N_{L} - B_L$ and
using the upper bound in Equation~\ref{eq:asdf1} with the lower bound
in Equation~\ref{eq:asdf} we find:
\begin{equation}  
D - \epsilon^\prime \leq \frac{ ((N_L/L^d) - (B_L/L^d) ) v_d(r)}{(1 -
2r/L)^d} \leq \frac{D + \epsilon^\prime}{(1 - 2r/L)^d} -
\frac{(B_L/L^d)\, v_d(r)}{(1 - 2r/L)^d}
\end{equation}  
Solving for $B_L/L^d$ we find that:  
\begin{equation}  
\frac{B_L}{L^d} \leq  
\frac{D}{v_d(r)} \left[ -(1- 2r/L)^d + 1\right]  +  
\frac{\epsilon^\prime}{v_d(r)} \left[ 1 + (1 - 2r/L)^d\right]  
\end{equation}  
This tells us that $B_L/L^d$ can be made as small as desired by
picking sufficiently small $\epsilon^\prime$ and $\tau = r/L$.
Finally, let S be any finite union of squares $S_{i}$ of sides $L_i$
where every $L_i$ exceeds some given $L_0$ and the $S_i$ intersect at most
on their boundaries. Taking $N_i$ to be the number of spheres in $C_r$
intersecting $S_i$, with $B_i$ the number of spheres intersecting the
boundary, we have the following bound on the density of the cover of
of $S$:
\begin{equation}  
\label{eq:lllk}  
\frac{\sum_i  (N_i - B_i)/L_i^d \, v_d(r) \, L_i^d}{\sum_i L_i^d}  
\leq  
D(S,C_r)  
\leq  
\frac{\sum_i  N_i /L_i^d \, v_d(r) \, L_i^d}{\sum_i L_i^d}  
\end{equation}  
By picking $r/L_0$ to be small enough we can make $N_i \,
v_d(r)/L_i^d$ as close as we want to $D(R^d,C_r)$ and $B_i/L_i^d$ as
close as we want to zero.  Consequently, since all the sums in
Equation~\ref{eq:lllk} are finite we can see that for any choice of
$\epsilon > 0$, for sufficiently small $r/L_0$, $|D(S,C_r) -
D(R^d,C_r)| < \epsilon$.  This proves the lemma. $\Box$
  
Lemma~\ref{lemma:constr} has given us some understanding of covers of finite
unions of squares.  The next lemma gives control over covers of
arbitrary Lebesgue measurable subsets of $R^d$.  The basic difficulty
that we must confront is that there are subsets of $R^d$ of Lebesgue
measure zero for which the covering density is not well defined.
Since we are interested in integration on parameter manifolds it is
natural that such sets of measure zero will not contribute to the
integral over the manifold.  The following lemma shows how to find
well-behaved subsets of any Lebesgue measurable set.
\begin{lemma}  
\label{lemma:seq}  
Let $D(d)$ be the minimal density for covering $R^d$ by spheres.  Let
$C = \{C_{r_1},C_{r_2},\cdots\}$ be any sequence of covers of $R^d$
such that $r_i \rightarrow 0$ as $i \rightarrow \infty$ and
$D(R^d,C_{r_i}) = D(d)$ for every $i$.  Take $G \subset R^d$ to have a
finite Lebesgue measure.  Then there exists a sequence $H_k \subseteq
G$ such that a) $\lim_{k \rightarrow \infty} \mu(G - H_k) = 0 $ and b)
$\lim_{i\rightarrow \infty} D(H_k,C_{r_i}) = D(d)$.
\end{lemma}  
{\bf Proof: } Let $G \subset R^d$ have finite Lebesgue measure.  Let
$H = \overline{G^\circ}$ be the closure of the interior of $G$ which
differs from $G$ at most by a set of measure zero.  Then $H$ can be
written as a countable union of squares $S_i$ each of which has finite
measure and which intersect at most on their boundaries.  Let $R_k =
\{S_i : \mu(S_i) > 1/k^2 \}$ be the set of these squares that have
side greater than $1/k$.  It is clear that $H_k = \bigcup_{S_i \in
R_k} S_i \subseteq H$ and that $\lim_{k\rightarrow\infty} \mu( H -
H_k) = 0$.  This proves the first part of the lemma.  Each $H_k$ is a
finite union of squares of side greater than $1/k$ that intersect at
most on their boundaries.  So, by Lemma~\ref{lemma:constr}, for any
$\epsilon > 0$ there is a $\tau_0$ such that if $r_i k < \tau_0$, then
$|D(H_k,C_{r_i}) - D(d)| < \epsilon.$ Since $r_i \rightarrow 0$ in the
limit $i\rightarrow \infty$, $D(H_k,C_{r_i}) \rightarrow D(d)$.  This
proves the second part of the lemma. $\Box$

\newtheorem{definition}{Definition}  
  
We have found that in the limit that the radius of covering spheres
$r$ goes to zero, any subset of $R^d$ of finite Lebesgue measure can
be covered up to a set of measure zero with a minimal thickness
$D(d)$.  We will now use this lemma to construct a measure on a
parameter manifold that reflects indistinguishability and translation
invariance.  Define a {\it regular sequence} $\{H_k\}$ of a Lebesgue
measurable set $G$ to be one of the sequences $\{H_k\}$ whose
existence was shown in Lemma~\ref{lemma:seq}.  Now consider one of the
sets $K_p$ in which the function $f$ and the Fisher Information
$J_{ij}$ are constant.  By rescaling the coordinates of $K_p$ by
$J_{ij}$ we transform the volumes of indistinguishability into spheres
of volume:
\begin{equation}                                 
\label{eq:v1}                                 
\tilde{V}_{\epsilon,\beta^*,N}= (2\pi\kappa/N)^{(d/2)}/ \Gamma(d/2 +
1)
\end{equation}                                 
and change the measure of $K_{p}$ from $\mu(K_{p})$ to $\sqrt{\det J}
\, \mu(K_{p})$ where $\mu$ is the Lebesgue measure in the original
coordinates.  Now suppose that we want to integrate the step function
$f$ over the measurable domain $I$.  Let $K_{pI} = K_{p} \bigcap I$
and let $\{H_{pIk}\}$ be a regular sequence of $K_{pI}$.  The
transformed coordinates define an embedding of $K_{pI}$ into $R^d$ and
we consider a minimal covering of $R^d$ by transformed volumes of
indistinguishability $\tilde{V}_{\epsilon,\beta^*,N}$. This minimal
covering induces a cover of $K_p$ and therefore of each $H_{pIk}$.  We
define a measure $\nu_{\epsilon\beta^*Nk}$ at levels $\epsilon,
\beta^*, N$ and $k$ for integration over $I$ by placing a delta
function at some point in the intersection of each covering sphere and
$H_{pIk}$. This yields the
following definition of integration of the step function $f$:
\begin{definition}
\label{def:def1}
Let $\{K_p\}$ be the sets on which the step maps $f$ and $J_{ij}$ are
both constant.  Then, at levels of indistinguishability $\epsilon$,
$\beta^*$ and $N$, and at level $k$ in a regular sequence of each
$K_p$, we define the integral of $f$ over the measurable domain $I$ to
be: 
\begin{equation}                    
\label{eq:int1}                 
\int_I f \, d\nu_{\epsilon\beta^*Nk} = \sum_p f_p\, N_{pIk}                 
\end{equation}      
where $N_{pIk}$ is the number of spheres that intersect $H_{pIk}
\subseteq K_p \subset R^d$ in the cover of $R^d$ by the
spheres $\tilde{V}_{\epsilon,\beta^*,N}$.
\end{definition}
We are actually interested in a measure
$\gamma_{\epsilon\beta^*Nk}$ normalized so that the integral of $1$
over the entire manifold gives unity.  The normalization is easily
achieved by dividing Equation~\ref{eq:int1} by the integral of $1$
over the manifold ${\cal M}$.
\begin{definition}
\label{def:def2}
Let $\{H_{pk}\}$ be a regular sequence of $K_p$ and let $N_{pk}$ be the
number of spheres that intersect $H_{pk} \subseteq K_p \subset R^d$
in the cover of $R^d$ by the spheres
$\tilde{V}_{\epsilon,\beta^*,N}$. The normalized integral of the step
function $f$ over the domain $I$ is given by:
\begin{equation}                 
\label{eq:int2}                 
\int_I f \, d\gamma_{\epsilon\beta^*Nk} = 
\frac{\sum_p f_p \,   N_{pIk}/N^{d/2}}
     {\sum_p  \,   N_{pk}/N^{d/2}}                                                  
\end{equation}                 
\end{definition}
The division by $N^{d/2}$ is motivated by our desire to take the
limit $ N \rightarrow \infty$.  The definition in
Equation~\ref{eq:int2} reflects the properties of indistinguishability
and translation invariance by ignoring variations on a scale smaller
than the volumes $V_{\epsilon,\beta^*,N}$ and giving equal weight in
the integral to all such volumes.
                 
      We begin by taking the limit $N \rightarrow \infty$ so that the
definition of the integral reflects indistinguishability in the limit
of an infinite amount of data.  This is followed by the limit $k
\rightarrow \infty$ so that the entire domains $K_{pI}$ are included
in the integral up to a set of measure zero.   Then we will take the
limits $\beta^* \rightarrow 1$ so that we are working with truly
indistinguishable distributions. Finally we will take $L^1$
completions of $f$ and $J_{ij}$ to arrive at the defintion of
integration of any Lebesgue measurable $f$ over a parmeter manifold
with Lebesgue measurable and non-singular Fisher Information.  The
result of this sequence of limits is summarized in the following
theorem. 
\begin{theorem}
\label{th:measure}
Let $\mu$ be the Lebesgue measure on a parameter manifold ${\cal M}$
that is induced by the parametrization.  Let $f$ be any Lebesgue
measurable function on the manifold and let the Fisher Information
$J_{ij}$ be Lebesgue measurable and non-singular everywhere on the
manifold.  Let $\gamma$ be the normalized measure on ${\cal M}$ that
measures the volume of distinguishable distributions indexed by the
parameters. Then $\gamma$ is absolutely continuous with respect to
$\mu$ and if $I$ is any Lebesgue measurable set, then:
\begin{equation}
\int_I f\,d\gamma = \frac{\int_I f\, \sqrt{\det{J_{ij}}} \, d\mu}
                         {\int \sqrt{\det{J_{ij}}} \, d\mu}
\end{equation}
\end{theorem}  
{\bf Proof:} First of all, observe that the volumes
$\tilde{V}_{\epsilon,\beta^*,N}$ used in covering the $H_{pIk}$ have
radius $r = \kappa/N$ .  Consequently, the sequence of covers by
$\tilde{V}_{\epsilon,\beta^*,N}$ for increasing $N$ and the sets of
the regular sequence $\{ H_{pIk} \} $ satisfy the conditions of
Lemma~\ref{lemma:seq}.  Therefore, applying the lemma and the
definition of the density of a cover (Equation~\ref{eq:dendef}), 
we find:
\begin{eqnarray}                                 
\lim_{k\rightarrow\infty} \lim_{N\rightarrow\infty}N_{pIk}/ N^{d/2}
&=& \lim_{k\rightarrow\infty} D(d) \mu(H_{pIk}) \sqrt{\det{J_p}}
\frac{\Gamma(d/2 +1)}{ (2\pi\kappa)^{d/2}} \nonumber \\ & = & D(d)
\mu(K_{pI}) \sqrt{\det{J_p}} \frac{\Gamma(d/2 +1)}{
(2\pi\kappa)^{d/2}}
\end{eqnarray}                 
where $J_p$ is the Fisher Information in the region $K_p$. 
(We have used the fact that the measure of $K_{pI}$ in the
coordinates in which the volumes of indistinguishability are spheres
is $\mu(K_{pI}) \sqrt{\det{J_p}}$.) Therefore,
both the numerator and denominator of the right hand side of
Equation~\ref{eq:int2} are finite sums of terms that approach finite
limits as $N\rightarrow\infty$.  So we can evaluate the limits of
these terms to write:
\begin{eqnarray}                                 
\lim_{k\rightarrow\infty} \lim_{N \rightarrow \infty} \int_I f \,                 
d\gamma_{\epsilon\beta^*Nk} & = &                                 
\frac{ \sum_p f_p     \,                            
D(d) \, \mu(K_{pI}) \sqrt{\det{J_p}} \Gamma(d/2 + 1)                                 
 / (2\pi\kappa)^{d/2} }                                 
       {  \sum_p  \,                                 
D(d) \, \mu(K_{p}) \sqrt{\det{J_p}} \Gamma(d/2 + 1)                                 
 / (2\pi\kappa)^{d/2} } \nonumber \\                                 
& = &                                 
\frac{\sum_p f_p \sqrt{\det{J_p}} \: \mu(K_{pI})}{\sum_p
\sqrt{\det{J_p}} \: \mu(K_{p})}
\end{eqnarray}                                 

The right hand side is now independent of $\beta^*$ and $\epsilon$
permitting us to freely take the limits $\beta^* \rightarrow 1$ and
$\epsilon \rightarrow 0$ which gives us the definition of a normalized
integral over truly indistinguishable distributions which we write as
$\int f d\gamma$.
                 
    We now want to take the $L^1$ completion of the step maps $J_{ij}$
and $f$ in order to arrive at the definition of integration of any
function that is Lebesgue measurable on the manifold.  First we take
the $L^1$ completion of the $J_{ij}$ with respect to Lebesgue measure.
By the standard theory of integration, the sums $\sum_p
\sqrt{\det{J_p}} \, \mu(K_{pI})$ converge to integrals to give the
following definition of the integrals of step maps $f$.(\cite{lang})
\begin{equation}                                 
\int_I f d\gamma = \frac{\sum_i f_i \int_{A_i\cap I} \sqrt{\det{J}} d\mu}
                                 {\int \sqrt{\det{J}} d\mu}
\end{equation}                          
where the step function $f$ is constant on the sets $A_i$.   We have
arrived at a new measure on the manifold $\gamma(K) = (\int_K
\sqrt{\det{J}} d\mu) / \int \sqrt{\det{J}} d\mu$ where $\mu$ is the
original Lebesgue measure.  Since $\gamma$ and $\mu$ are absolutely
continuous with respect to each other, the $L^1$ completion of step
maps with respect to $\gamma$ describes the same class of the
functions as the completion of step maps with respect to $\mu$.  We
can therefore take the $L^1$ completion of $f$ with respect to
$\gamma$ to arrive at the following definition of integration of any
Lebesgue measurable function on a parameter manifold:
\begin{equation}                                 
\int_I f d\gamma = \frac{\int_I f \sqrt{\det{J}} d\mu}{\int
\sqrt{\det{J}} d\mu}
\end{equation}                                  
where $\mu$ is the Lebesgue measure induced by the parametrization.
As discussed above, this construction accounts for
indistinguishability and translation invariance in the space of
probability distributions. $\Box$

In sum, the normalized measure on the manifold that accounts for the
indistinguishability of neighbouring distributions is given by:
\begin{equation}                                 
\label{eq:appmeas}                                 
d\gamma = \frac{d\mu \, \sqrt{\det{J_{ij}}} } {\int \, d\mu \,
                  \sqrt{\det{J_{ij}}}} \label{eq:measure2}
\end{equation}                                 
where $d\mu$ is the Lebesgue measure on the manifold induced by its
atlas which in simple cases is simply the product measure $d^d\Theta =
\prod_{i=1}^{d} d\theta_i$.  In this expression we have taken the
limits $\beta^* \rightarrow 1$ and $N \rightarrow \infty$.  The
meaning of this is that we are dividing out the volume of the
parameter space which contains models that will be perfectly
indistinguishable (in a one-sided error) given an arbitrary amount of
data.  As discussed earlier, Equation~\ref{eq:appmeas} is equivalent
to a choice of Jeffreys prior in the Bayesian formulation of model
inference.  We stated earlier that Jeffreys prior has the desirable
property of being reparametrization invariant on account of the
transformation properties of the Fisher Information that enters its
definition, but that one could define many such
quantities.\footnote{Indeed Jeffreys considers priors related to
various distances such as the $L^k$ norms.(\cite{jeffreys})} It
appears that the derivation in this paper may provide the first
rigorous justification for a choice of Jeffreys prior for Bayesian
inference that does not involve assumption of a Minimum Description
Length principle or a statement concerning compact coding of data.
The derivation suggests that the Fisher Information is the
reparametrization invariant prior on the parameter manifold that is
induced by a uniform prior in the space of distributions.  As such it
would seem to be the natural prior for density estimation in a
Bayesian context.  It is worthwhile to point out that the work of
Wallace and Freeman and Barron and Cover (among others) has
demonstrated that the optimal code derived from a parametric model
should pick parameters from a grid distributed with a density
inversely proportional to the determinant of the Fisher Information
matrix.  (\cite{wallace},\cite{barron}) The continuum limit of these
grids can be obtained in the fashion demonstrated here and would yield
a Jeffreys prior on the parameter manifold.
                                 
The reader may worry that the asymmetric errors $\alpha \rightarrow 0$
and $\beta \rightarrow 1$ are a little peculiar since they imply that
$p$ can be distinguished from $q$, but $q$ cannot be distinguished
from $p$.  A more symmetric analysis can be carried out in terms of
the Chernoff bound at the expense of a convexity assumption on the
parameter manifold.  Since the derivation of the measure using the
Chernoff bound exactly parallels the derivation using Stein's Lemma
and yields the same result, we will not present it here.  Although the
derivation in this section has focussed on deriving the measure on a
parameter manifold, future sections will take the metric on the
manifold to be the Fisher Information.

\subsection{Riemannian Geometry in The Space of Distributions}                                 
I do not have enough space in this paper to recapitulate the theory of
Riemannian geometry in the setting of the space of probability
distributions.  I will therefore assume that the reader has a
rudimentary understanding of the notions of vectors, connection
coefficients and covariant derivatives on manifolds.  The necessary
background can be gleaned from the early pages of any differential
geometry or general relativity textbook.  A discussion of geometry in
a specifically statistical setting can be found in the works of Amari,
Rao and others.(\cite{amari85}, \cite{amari87}) In the next section I
will assume a rudimentary knowledge of geometry, but since we do not
need any sophisticated results, the reader who is unfamiliar with
covariant derivatives should still be able to understand most of the
results.  Table~\ref{table1} provides a few formulae that will be
useful to such readers, but contains no explanations.

\begin{table}                                
\label{table1}                                
\begin{center}                                
\fbox{                                 
\parbox{5.5 in}{                                 
{\it Vectors} are objects in the tangent space of a manifold.  We write them                                 
with {\it upper} indices as $V^\mu$.   {One forms} are duals to vectors.  We                                 
write them {\it lower} indices as $W_\mu$.   {\it Tensors} are formed                                 
by taking tensor products of vectors and forms.   The {\it metric} is                                 
a {\it rank 2 tensor} with two lower indices and is written as                                 
$g_{uv}$.  The {\it inverse metric} is written with upper indices as                                 
$g^{\mu\nu}$.                                  
In this paper the metric on a parameter manifold is found                                 
to be the Fisher Information $J_{\mu\nu}$.   We map between vectors and                                 
one-forms (between upper and lower indices) using the metric or its                                 
inverse: e.g.,                                  
$V_\mu = J_{\mu\nu}\,V^{\nu}$ and                                 
${K^\alpha_{\hspace{0.1in}\beta}}^{\gamma} =                                  
J_{\beta\mu} K^{\alpha\mu\gamma}$ where we use the {\it summation                                 
convention} that repeated indices are summed over.  We define                                 
the {\it covariant derivative} $D$  on a manifold which acts as follows                                 
on functions ($f$), vectors ($V^{\mu}$) and one-forms ($V_{\mu}$):                                 
\begin{eqnarray}                                 
D_{\mu} f &=& \partial_\mu f \nonumber \\                                 
D_{\mu} V^\alpha &=& \partial_\mu V^\alpha +                                 
\Gamma^\alpha_{\mu\nu}\,V^\nu \nonumber \\                                 
D_{\mu} V_\alpha &=& \partial_\mu V^\alpha -                                 
\Gamma^\nu_{\mu\alpha}\,V_\nu \nonumber                                 
\end{eqnarray}                                 
In these equations $\partial_\mu$ is the usual partial derivative with                                 
respect to the coordinate $\theta^\mu$.   Derivatives of higher tensors                                 
are defined analogously.  The $\Gamma^{\alpha}_{\beta\gamma}$ are the                                 
unique metric-compatible {\it  connection coefficients} defined as follows:                                 
\begin{equation}                                 
\Gamma^{\alpha}_{\beta\gamma} = \frac{1}{2}\, g^{\alpha\sigma}                                 
[\partial_\beta g_{\gamma\sigma} + \partial_\gamma g_{\beta\sigma}                                 
-\partial_\sigma g_{\beta\gamma}  ]                                  
\end{equation}                                 
The covariant derivative of the metric vanishes using this
connection and this elementary fact ia used in this paper. The
{\it curvature} of the manifold is measured by the failure of the
covariant derivative to commute. The characterization of curvature in
a statistical setting can be found in the work of Amari, Rao and
others.(\cite{amari85}, \cite{amari87}) }}
\end{center}                                 
\caption{Useful Geometric Equations}                                 
\end{table}

\section{Parsimony and Consistency of The Razor}                                 
\label{sec:largen}                                 
In the previous sections we have constructed the razor from Bayes'
Rule and discussed measures and metrics on parameter manifolds.  We
are left with a candidate for a coordinate invariant index of
simplicity and accuracy of a parametric family as a model of a
true distribution ~\true:
\begin{equation}                                 
R_N(A) = \frac{ \int d\mu(\Theta) \sqrt{\det{J_{ij}}} e^{-N D(\true \| \Theta) }}                                 
              { \int d\mu(\Theta) \sqrt{\det{J_{ij}}}}                                 
\end{equation}                                 
where the Fisher Information $J_{ij}$ is the metric on the manifold.
In this section I will demonstrate that the razor has the desired
properties of measuring simplicity and accuracy.  In order to make
progress various technical assumptions are necessary.
\mathcom{\best}{0}{\Theta^*} Let~\best~ be the value of $\Theta$ that
globally minimizes $D(t\|\Theta)$.  I will assume that~\best~is a
unique global minimum and that it lies in the interior of the compact
parameter manifold.  I will also assume that that $D(t\|\Theta)$ and
$J_{ij}(\Theta)$ are smooth functions of $\Theta$ in order that Taylor
expansions of these quantities are possible.  (Actually the degree of
continuity required here depends on the accuracy of the approximation
we seek and since we will only evaluate terms to $O(1/N)$ we will only
require the existence of derivatives up to the fourth order for our
computations.)  Finally, let the values of the local minima be bounded
away from the global minimum by some $b$.  For any given $b$, for
sufficiently large $N$, the value of the razor will be dominated by
the neighbourhood of~\best.  Our strategy for evaluating the razor
will be to Taylor expand the exponent in the integrand
around~\best~and to develop a perturbation expansion in powers of
$1/N$.  We will omit mention of the $O(\exp{-bN})$ terms arising from
the local minima.  In their analysis of the asymptotics of the
Bayesian marginal density Clarke and Barron introduce a notion of
``soundness of parametrization''.(\cite{clarke})\footnote{A parametric
family is said to be ``sound'' if convergence of a sequence of
parameter values is equivalent to weak convergence of the distributions
indexed by the parameters.}  This condition is intended to guarantee
that there is a one-to-one map between parameters and distributions
and that distant parameters index distant distributions.  In geometric
terms this simply means that the parameter manifold is embedded in the
space of distributions in such a way that no two separable points on
the manifold are embedded inseparably in the space of distributions -
i.e., the manifold does not fold back on itself or intersect itself.
The conditions stated for the following analysis are much weaker
because we only need ``soundness'' at the point on the manifold that
is closest to the true distribution in relative entropy.  Even this is
merely a technical condition for ease of analysis - multiple global
maxima of the integrand of the razor would simply contribute
separately to the analysis and thereby increase the value of the
razor.  The most important conditions required in this paper are that
Taylor expansions of the relevant quantities should exist at $\best$.

\mathcom{\fish}{0}{\tilde{J}}                                 
\mathcom{\grad}{0}{\nabla}                                 
                                 
\subsection{A Perturbative Expansion of The Razor}                                 
We can begin the evaluation of the razor by rewriting it as:                                 
\begin{equation}                                 
\label{eq:step1}                                 
R_N(A) = \frac{\int d\mu(\Theta) e^{[(1/2)Tr \ln J - N D(t \| \theta) ]}}                                 
              {\int d\mu(\Theta) \sqrt{\det J_{ij}}}                                 
\end{equation}                                 
where $Tr$ denotes trace. Define $F(\Theta) = Tr \ln J_{ij}$.  Let
$\fish_{\mu_1 \cdots \mu_i} = \grad_{\mu_1} \cdots \grad_{\mu_i}
D(t\|\Theta) |_{\Theta^*}$ be the nth covariant derivative of the relative
entropy with respect to $\theta^{\mu_1} \cdots \theta^{\mu_i}$
evaluated at~\best.  Define the nth covariant derivatives of
$F(\Theta)$ similarly.  We can Taylor expand the exponent in
Equation~\ref{eq:step1} in terms of these quantities.  Letting $E$ be
the exponent, we find that:
\begin{equation}                                  
\label{eq:expon}                              
E = - N \left[ D(t\|\best) +                                     
          \sum_{i=2}^{\infty} \frac{1}{i!}                                 
             \fish_{\mu_1\cdots\mu_i} \delta\theta^{\mu_1}                                 
      \cdots\delta\theta^{\mu_i} \right]  +                                    
   \frac{1}{2} F(\best) +  \sum_{i=1}^{\infty} \frac{1}{2 i!}                                   
F_{\mu_1\cdots\mu_i} \delta\Theta^{\mu_1}\cdots\delta\Theta^{\mu_i}                                 
\end{equation}                                 
To proceed further shift the integration variable $\Theta$ to $\Theta
- \best = \delta\Theta$ and rescale to integrate with respect to $\Phi
= \sqrt{N} \delta\Theta$.  With this change of variables the razor
becomes:
\begin{equation}                                 
R_N(A) = \frac{e^{-(N D(t\|\best) - \frac{1}{2} F(\best))} N^{-d/2}                 \int                                 
d\mu(\Phi) e^{-((1/2)\fish_{\mu_1\mu_2}\phi^{\mu_1}\phi^{\mu_2}                                                                     
+ G(\Phi))}}              {\int d\mu(\Theta) \sqrt{\det{J_{ij}}}}                                 
\end{equation}                                 
where $G(\Phi)$ collects the terms in the exponent that are suppressed
by powers of $N$:
\begin{eqnarray}                                 
\label{eq:gdef}                              
G(\Phi) = & \sum_{i=1}^{\infty} \frac{1}{\sqrt{N^i}} \left[                                 
                  \frac{1}{(i+2)!} \tilde{J}_{\mu_1\cdots\mu_{i+2}}                                 
                                   \phi^{\mu_1}\cdots\phi^{\mu_{(i+2)}}                                 
                  - \frac{1}{2i!} F_{\mu_1\cdots\mu_i}                                 
                             \phi^{\mu_1} \cdots \phi^{\mu_i}                                 
                        \right] \nonumber \\                                 
= & \frac{1}{\sqrt{N}} \left[ \frac{1}{3!} \tilde{J}_{\mu_1\mu_2\mu_3}                                 
\phi^{\mu_1} \phi^{\mu_2} \phi^{\mu_3} - \frac{1}{2} F_{\mu_1}                                 
\phi^{\mu_1} \right] + \nonumber \\                                  
   &        \frac{1}{N}\left[ \frac{1}{4!} \tilde{J}_{\mu_1\cdots\mu_4}                                 
\phi^{\mu_1}\cdots\phi^{\mu_4} - \frac{1}{2\,2!} F_{\mu_1\mu_2} \phi^{\mu_1}                                 
\phi^{\mu_2} \right] +                                    
           O(\frac{1}{N^{3/2}})                                 
\end{eqnarray}                                 
Note that the leading term in $G(\Phi)$ is $O(1/\sqrt{N})$.  The razor
may now be evaluated in a series expansion using a standard trick from
statistical mechanics.  Define a ``source'' $h = \{h_1 \ldots h_d\}$
as an auxiliary variable.  Then it is easy to verify that the razor
can be written as:
\begin{equation}                              
\label{eq:gdef2}                                 
R_N(A) = \left. \frac{e^{-(N D(t\|\best) - \frac{1}{2} F(\best))}                                   
         e^{-G(\grad_h)}                                           
      \int d\mu(\Phi) e^{-(\frac{1}{2} \fish_{\mu_1\mu_2}\phi^{\mu_1}                                 
                                                 \phi^{\mu_2}                                  
                                     +  h_\mu \phi^\mu)}}                                 
              {N^{d/2}\; \int d\mu(\Theta) \sqrt{\det{J_{ij}}}} \right|_{h=0}                                    
\end{equation}                                 
where the derivatives have been assumed to commute with the integral.
The function $G(\Phi)$ has been removed from the integral and its
argument ($\Phi = (\phi^1\ldots\phi^d)$) has been replaced by $\grad_h
= \{\partial_{h_1} \ldots \partial_{h_d} \} $.  Evaluating the
derivatives and setting $h = 0$ reproduces the original expression for
the razor.  But now the integral is a simple Gaussian.  The only
further obstruction to doing the integral is that the parameter space
is compact and consequently the integral is a complicated
multi-dimensional error function.  As our final simplifying assumption
we will analyze a situation where~\best~is sufficiently in the
interior, or $N$ is sufficiently large as to give negligible error
when the integration bounds are extended to infinity.  The integral
can now be done instantly.  We find that:
\begin{equation}                                 
\label{eq:temp2}                                 
R_N(A) = \frac{e^{-(N D(t\|\best) - (1/2) F(\best))}                                 
                e^{-G(\grad_h)}                                  
              \left[ \left( \frac{(2\pi)^d}{\det\fish} \right)^{1/2}                                  
                   \exp{-(h^{\mu_1} \fish^{-1}_{\mu_1\mu_2} h^{\mu_2})}                                 
               \right]_{h=0}}                                 
              {N^{d/2}\; \int d\mu(\Theta) \sqrt{\det{J_{ij}}}}                                                 
\end{equation}                                 
Expanding $\exp G$ and collecting terms gives:                                 
\begin{equation}                                 
R_N(A) = \frac{1}{V} e^{-N D(t\|\best)}                                 
\left(\frac{2\pi}{N}\right)^{d/2}                                 
\left(\frac{\det{J_{ij}(\Theta^*)}}{\det{\tilde{J_{\mu\nu}}}}\right)^{1/2} \left[ 1 +                 
O(\frac{1}{N})                                 
\right]                                   
\end{equation}                                 
where we have defined $V= \int d\mu(\Theta) \sqrt{\det{J}}$ to be the volume of                                 
the parameter  manifold measured in the Fisher Information                                 
metric.\footnote{The terms of order $1/\sqrt{N}$ integrate to zero                                 
because they are odd in $\phi$ while the Gaussian integrand and the                                 
integration domain in our approximation are even in $\phi$.}  The                                 
terms of order $1/N$ arise from the action of  $G$ in                                 
Equation~\ref{eq:temp2}.   It turns out to be most useful to examine                                 
$\chi_N(A) \equiv - \ln R_N(A)$.  In that                                 
case, we can write, to order $1/N$:                                 
\begin{eqnarray}                                 
\label{eq:lograzor}                                 
\chi_N(A) = & N D(t \| \Theta^*) + \frac{d}{2} \ln{N}                                 
- \frac{1}{2} \ln{(\det{J_{ij}(\Theta^*)}/\det{\tilde{J}_{\mu\nu}})} -                                 
\ln{\left[\frac{(2\pi)^{d/2}}{V}\right]} + \nonumber \\                                 
 & \frac{1}{N}  \left\{ \frac{\fish_{\mu_1\mu_2\mu_3\mu_4}}{4!} \left[                                 
(\fish^{-1})^{\mu_1\mu_2} (\fish^{-1})^{\mu_3\mu_4} + \ldots \right] -                                 
     \frac{F_{\mu_1\mu_2}}{2\,2!} \left[ (\fish^{-1})^{\mu_1\mu_2} +                                 
(\fish^{-1})^{\mu_2\mu_1}                                 
\right]                                  
\right. - \nonumber \\                                 
 &    \frac{\tilde{J}_{\mu_1\mu_2\mu_3} \fish_{\nu_1\nu_2\nu_3}}{2!\,3!\,3!}                                 
 \left[                                 
(\fish^{-1})^{\mu_1\mu_2}(\fish^{-1})^{\mu_3\nu_1}(\fish^{-1})^{\nu_2\nu_3}                                 
+ \ldots \right] - \nonumber \\                                 
 &  \left. \frac{F_{\mu_1} F_{\mu_2}}{2!\,4\,2!\,2!} \left[                                 
(\fish^{-1})^{\mu_1\mu_2} + \ldots                                  
\right]  + \frac{F_{\mu_1} \fish_{\mu_2\mu_3\mu_4}}{2!\,2\,2!\,3!}                                 
\left[(\fish^{-1})^{\mu_1\mu_2} (\fish^{-1})^{\mu_3\mu_4} + \ldots                                 
\right]                                  
\right\}                               
\end{eqnarray}                                 
The ellipses within the parentheses indicate further terms involving all                                 
permutations of the indices on the single term that has been indicated and we have                 
omitted terms of $O(1/N^2)$ and smaller.   It is worthwhile to point out that the                 
systematic series expansion  above allows us to evaluate the razor to arbitrary accuracy                 
for any                                 
relative entropy functions whose Taylor expansion exists and whose                                 
derivatives grow sufficiently slowly with order.   Therefore, this method of                                 
analyzing the asymptotics circumvents the need to place bounds on the                                 
higher terms since they can be explicitly evaluated.   The statistical                                 
mechanical idea of using                                 
such expansions around a saddlepoint of an integral could also find                                 
applications in other asymptotic analyses in information theory                                  
in which integrals are dominated by narrow maxima.\footnote{A simple                                 
form of this method of integration has appeared before under the                                 
rubric ``Laplace's Method'' in the work of Barron and others.(\cite{barron85})}                                 
In the next section we will discuss why this large $N$ analysis shows                                 
that the razor meaures simplicity and accuracy and we will analyze the                                 
geometric meaning of the terms in the above expansion.   We will then discuss                                 
the connections between Equation~\ref{eq:lograzor} and the                                 
Minimum Description Length Principle.                                 
                                 
\subsection{Parsimony and Consistency}                                 
The various terms of Equation~\ref{eq:lograzor} tell us why the razor                                 
is a measure of the simplicity and accuracy of a parametric                                 
distribution.   Models with higher values of $R_N(A)$ and therefore                                 
lower values of $\chi_N(A)$ are considered to be better.  The $O(N)$                                 
term, $N D(t||\best)$ measures the relative                                  
entropy  between the true distribution and the best model                                 
distribution on the manifold.   This is a measure of the accuracy with                                 
which the model family $A$ will be able to describe $t$.  The geometric                                 
interpretation of this term is that it arises from the distance between                                 
the true distribution and the closest point on the model manifold in                                 
relative entropy sense.  The $O(\ln N)$                                 
term, $(d/2) \ln N$,  tells us that the value of the log razor increases                                 
linearly in the dimension of the parameter space.  This penalizes                                 
models with many degrees of freedom.  The geometric reason for the                                 
existence of this term is that the volume of a peak in the integrand                                 
of the razor measured relative to the volume of the manifold shrinks                                 
more rapidly as a function of $N$ in higher dimensions.                                 
 The $O(1)$ term, is even more                                 
interesting.    The determinant of $\fish_{ij}^{-1}$ is proportional to                                 
the volume of the ellipsoid in parameter space around $\Theta^*$ where                                 
the value of the                                  
integrand of the razor is significant.\footnote{If we fix a fraction                                 
$f < 1$ where $f$ is close to 1, the integrand of the razor will be                                 
greater that $f$ times the peak value in an elliptical region around                                 
the maximum.}     The scale for determining whether $\det                                 
\fish_{ij}^{-1}$ is large or small is set by the Fisher Information on the                                 
surface whose determinant defines the volume element.  Consequently                                 
the term $(\det J/\det\fish)^{1/2}$ can be understood as measuring the                                 
naturalness of the model in the sense discussed in                                 
Section~\ref{sec:qual} since it involves a preference for model                                 
families with distributions concentrated around the true.                                 
Another way of understanding this point is to observe from the                                 
derivation of the integration measure in the razor that given a fixed                                 
number of data points $N$, the volume of indistinguishability around                                 
$\Theta^*$ is proportional to $(\det{J})^{-1/2}$.  So the factor                                 
$(\det{J}/\det{\fish})^{(1/2)} $                                 
is essentially proportional to the   ratio $V_{large}/V_{indist}$, the                                 
ratio of the volume where the integrand of the razor is large to the                                 
volume of indistinguishability introduced earlier.   Essentially, a                                 
model is better (more natural) if there are many distinguishable                                 
models that are close to the true. The term                                  
$\ln{(2\pi)^d/V}$ can be understood as a preference for models that                                 
have a smaller invariant volume in the space of distributions and                                 
hence are more constrained.  The terms proportional to $1/N$ are less                                 
easy to interpret.   They involve higher derivatives of the metric on                                 
the parameter manifold and of the relative entropy distances between                                 
points on the manifold and the true distribution.   This suggests that                                 
these terms essentially penalize high curvatures of the model                                 
manifold, but it is hard to extract such an interpretation in terms of components of the                 
curvature tensor on the manifold.                   
                                 
    A consistent estimator of the razor can be used to implement                                 
parsimonious and consistent inference.  Suppose we are comparing two                                 
model families A and B.  We evaluate the razor of each family and pick                                 
the one with the larger razor.  To evaluate the behaviour of the razor                                 
we have consider several different cases.  First suppose that A is                                 
d-dimensional and B is a more accurate k-dimensional model with $k >                                 
d$.  By saying that B is more accurate we mean that $D(t\|\Theta^*_B)                                 
<D(t\|\Theta^*_A) $. We expect that for small $N$ the terms                                 
proportional to $\ln N$ will dominate and that for large $N$ the terms                                 
proportional to $N$ will dominate.  We can compute the crossover                                 
number of events beyond which accuracy is favoured over simplicity.                                 
Ignoring the terms of $O(1)$ let us ask how large $N$ must be so that                                 
$R_N(B) \geq R_N(A)$.    The answer is easily seen to be the solution                                 
to the equation:                                 
\begin{equation}                                 
\label{eq:cross1}                                 
\frac{(k - d)}{2} \frac{\ln{N}}{N} \leq \Delta D + O(1/N) =                                 
D(t\|\Theta^*_A) - D(t\|\Theta^*_B) + O(1/N)                                 
\end{equation}                                 
Up to terms of $O(1/N)$, this is  the expected                                 
crossover point between A and B if the inference used the Minimum                                 
Description Length principle based on stochastic complexity as introduced                                 
by Rissanen.(\cite{riss84},\cite{riss86})   The $O(1)$ terms in the                                 
razor are important for small $N$ and for cases where the models in                                 
question have parameter spaces of equal dimension.   In that case,                                 
ignoring terms of $O(1/N)$ in the razor, the crossover point where                                 
$R_N(B) \geq R_N(B)$ is given by:                                 
\begin{equation}                                 
N \geq \frac{1}{D(t\|\Theta^*_A) - D(t\|\Theta^*_B)}                                 
         \left[ \ln\left(\frac{V_B}{V_A}\right)                                 
             +\frac{1}{2} \ln\left( \frac{J_B}{\fish_B}                                  
                                     \frac{\fish_A}{J_A} \right) \right]                                 
\end{equation}                                 
The terms within the parentheses have been interpreted above in terms                                 
of the relative volume of the parameter space that is close to the true                                 
distribution (in other words, as a measure of robustness).   So we see                                 
that if A is a more robust model than B then the crossover number of                                 
events is greater.    The crossover point is inversely proportional to                                 
the difference in relative entopy distances between the true                                 
distribution and the best model and this too makes good intuitive sense.                                 
                                 
Further examinations of this sort show that the razor has a preference                                 
for simple models, but is consistent in that the most accurate model                                 
in relative entropy sense will dominate for sufficiently large N.                                 
Inference can be carried out with a countably large set of candidate                                 
families by placing the families in a list and examining the razor of                                 
the first $N$ families when $N$ events are provided.  It is clear that                                 
this procedure is then guaranteed to be asymptotically consistent                                 
while remaining parsimonious at each stage of the inference.  Indeed                                 
the model which is closest to the true in relative entropy sense will                                 
eventually be chosen while simpler models may be be preferred for                                 
finite $N$.                                 
                                 
\section{Various Meanings of The Results}                                 
\label{sec:meaning}                                
\subsection{Relationship to The Asymptotics of Bayes Rule}                    
\label{sec:relbayes}                                
The analysis of the previous section has shown that the razor is
parsimonious, yet consistent in its preferences.  Unfortunately, in
order to compute the razor one must already know the true
distribution.  It is certainly an index measuring the simplicity and
accuracy of a model, but actual inference procedures must devise
schemes to {\it estimate} the value of the razor from data.  A good
estimator of the razor will be guaranteed to pick accurate, yet simple
models.  So how do we estimate the razor of a model?
                                 
Given the Bayesian derivation of the razor a natural candidate is the
Bayesian posterior probability of a parametric model given the data.
As discussed before, this probability is given by:
\begin{equation}                                 
\label{eq:razest}                                 
R_E(A) = \frac{ \int d\mu(\Theta) \sqrt{\det{J}}
                          \exp(\ln{\Pr(E|\Theta))}} { \int
                          d\mu(\Theta) \sqrt{\det{J}}}
\end{equation}                                 
We have used a Jeffreys prior which is the uniform prior on the space
of probability distributions as discussed in previous sections.  The
relationship between the razor and the estimator in
Equation~\ref{eq:razest} can be analyzed in various ways.  The
simplest relationship arises because the exponential is a convex
function so that Jensen's Inequality gives us the following bound on
the expectation value of $R_E(A)$ in the true distribution $t$:
\begin{equation}                                 
\label{eq:lower}                                 
<R_E(A)>_t \, \, \geq \, \, \frac{\int d\mu(\Theta) \sqrt{\det{J}}
\exp{<\ln \Pr(E|\Theta)>_t}} {\int d\mu(\Theta) \sqrt{\det{J}}} \, =
\, R_N(A) \, e^{- N h(t)}
\end{equation}                                 
where $h$ is the differential entropy of the true distribution.  So
the razor times the exponential of the entropy of the true is a lower
bound on the expected value of the Bayesian posterior.
                 
    A sharper analysis may be carried out to show that under certain
regularity assumptions the razor reflects the typical behaviour of
$\chi_E(A) \equiv -\ln(R_E(A)) - N h(t)$.  The first assumption is
that $\ln{\Pr(E|\Theta)}$ is a smooth function of $\Theta$ for every
set of outcomes $E =\{e_1, \cdots e_N\}$ in the $N$ outcome sample
space.  (In fact, this assumption can be weakened to smoothness only
in a neighbourhood of $\Theta^*$.)  Using this premise, and the
already assumed smoothness of Fisher Information matrix
$J_{ij}(\Theta)$, we can expand the exponent in
Equation~\ref{eq:razest} around the maximum likelihood parameter
$\hat{\Theta} = \arg\max_\Theta \ln{\Pr(E|\Theta)}$ to obtain:
\begin{eqnarray}                                 
\label{eq:estexp}                                 
E & = & - N\left[ - \frac{\ln\Pr(E|\hat{\Theta})}{N} +
                            \sum_{i=2}^{\infty} \frac{1}{i!}
                            \tilde{I}_{\mu_1\cdots\mu_i}|_{\hat{\Theta}}\,
                            \delta\Theta^{\mu_1}\cdots
                            \delta\Theta^{\mu_i} \right] \nonumber \\
                            & & + \frac{1}{2} F(\Theta^*) +
                            \sum_{i=1}^{\infty} \frac{1}{2i!} \,
                            F_{\mu_1 \cdots \mu_i} \,
                            \delta\Theta^{\mu_1}\cdots\delta\Theta^{\mu_i}
\end{eqnarray}                                 
In this expression $F(\Theta^*)$ and $F_{\mu_1\cdots\mu_i}$ are the
same as in Equation~\ref{eq:expon} and we have defined
$\tilde{I}_{\mu_1\cdots\mu_i} = - \grad_{\mu_1} \cdots \grad_{\mu_i}
\ln\Pr(E|\Theta) /N$.
We will only consider models in which $\tilde{I}_{\mu\nu}$, the
empirical Fisher Information related to relative entropy distances
between model distributions and the true, is nonsingular everywhere
for every set of outcomes $E$.  This is a condition ensuring that
nearby parameters index sufficiently different models of the true
distribution.  By imitating the analysis of the razor (under the same
assumptions as those listed for that analysis), we find that:
\begin{equation}                              
R_E(A) = \frac{e^{\ln\Pr(E|\hat{\Theta}) + \frac{1}{2}
F(\hat{\Theta})} e^{- \tilde{G}(\grad_h)}
\left[\left(\frac{(2\pi)^d}{\det{\tilde{I}}} \right)^{1/2}
e^{-h^{\mu_1} \tilde{I}^{-1}_{\mu_1\mu_2}|_{\hat{\Theta}} h^{\mu_2} }
\right] }{N^{d/2} \; V }
\end{equation}                              
where $\tilde{G}$ is the same as in Equations~\ref{eq:gdef}
and~\ref{eq:gdef2} with the substitution of $\tilde{I}$ for every
$\tilde{J}$.  Defining $\chi_E(A) \equiv - \ln{R_E(A)} - N h(t)$ we
find that to $O(1/N)$:
\begin{eqnarray}                                   
\chi_E(A) = & N \left(\frac{-\ln\Pr(E|\hat{\Theta})}{N} - h(t) \right)
+ \frac{d}{2}\ln{N}
-\frac{1}{2}\ln(\det{J_{ij}(\hat{\Theta})}/\det{\tilde{I}_{\mu\nu}|_{\hat{\Theta}}})
\nonumber \\ & -\ln\left[\frac{(2\pi)^{d/2}}{V}\right] + + O(1/N)
\label{eq:chie}
\end{eqnarray}                              
Terms proportional to positive powers of $1/N$ may be computed as
before, but we will not evaluate them explicitly here.  It suffices to
note that all terms of order $1/N^k$ in Equation~\ref{eq:chie} are
identical to the corresponding terms in Equation~\ref{eq:lograzor}
with $\tilde{I}$ substituted for $\tilde{J}$.
                
We will now prove a theorem showing that any finite collection of
terms in $\chi_E(A)$ converges with high probability to $\chi_N(A) =
-\ln{R_N(A)}$ in the limit of a large number of samples.  Throughout
the discussion we will assume {\it consistency} of the maximum
likelihood estimator in the following sense.  Let $U$ be any
neighbourhood of $\Theta^* = \arg\min_\Theta D(t\|\Theta)$ on the
parameter manifold and let $E =\{e_1\cdots e_N\}$ be any set of $N$
outcomes drawn independently from $t$.  Then, for any $0<\delta<1$, we
shall assume that the maximum likelihood estimator $\hat{\Theta} =
\arg\max_{\Theta} \ln{\Pr(E|\Theta)}$ falls inside $U$ with
probability greater than $1 - \delta$ for sufficiently large N.  We
also require that the log likelihood of a single outcome $e_i$,
$\ln\Pr(e_i|\Theta)$, considered as a family of functions on $\Theta$
indexed by the outcomes $e_i$, is an {\it equicontinuous} family at
$\Theta^*$.(\cite{lang}) In other words, given any $\epsilon > 0$,
there is a neighbourhood $M$ of $\Theta^*$, such that for every $e_i$
and $\Theta \in M$, $|\ln\Pr(e_i|\Theta) - \ln\Pr(e_i|\Theta^*)| <
\epsilon$.  Finally, we will require that all derivatives with respect
to $\Theta$ of the log likelihood of a single outcome should be
equicontinuous at $\Theta^*$ in the same sense.
             
\begin{lemma}                
\label{lemma:conv1}                
Let $N$ be the number of iid outcomes $E=\{e_1\cdots e_N\}$ arising
from a distribution $t$ and take $\epsilon > 0 $ and $0 < \delta < 1$.
If the maximum likelihood estimator is consistent,
$Pr(|J_{ij}(\hat{\Theta}) - J_{ij}(\Theta^*)| > \epsilon) < \delta $
for sufficiently large $N$. (See above for definitions of $\Theta^*$
and $\hat{\Theta}$.)  Furthermore, if the log likehood of a single
outcome is equicontinuous at $\Theta^*$ (see definition above) then
$\Pr(|(-1/N) \ln\Pr(E|\hat{\Theta}) - D(t|\Theta^*) - h(t))| >
\epsilon) < \delta$ for sufficiently large $N$.  Finally, if the
derivatives with respect to $\Theta$ of the log likelihood of a single
outcome are equicontinuous at $\Theta^*$, then
$\Pr(|\tilde{I}_{\mu_1\cdots\mu_i}|_{\hat{\Theta}} -
\tilde{J}_{\mu_1\cdots\mu_i}| > \epsilon) < \delta$ for sufficiently
large $N$.
\end{lemma}                
{\bf Proof: } We have assumed that the Fisher Information matrix
$J_{ij}(\Theta)$ is a smooth matrix valued function on the parameter
manifold.  By consistency of the maximum likelihood estimator
$\hat{\Theta} \rightarrow \Theta^*$ in probability.  Since the entries
of the matrix $J_{ij}$ are continuous functions of $\Theta$ we can
conclude that $J_{ij}(\hat{\Theta}) \rightarrow J_{ij}(\Theta^*)$ in
probability also.  This proves the first claim. To prove the second
and third claims consider any function of the form $F_N(E,\Theta) =
(1/N) \sum_{i=1}^N F_1(e_i,\Theta)$ where $F_1(e_i,\Theta)$ is an
equicontinuous family of functions of $\Theta$ at $\Theta^*$.  We want
to show that $|F_N(E,\hat{\Theta}) - E_t[F_1(e_i,\Theta^*)]|$
approaches zero in probability where the expectation is taken in $t$,
the true distribution.  To this end we write:
\begin{equation}            
\label{eq:bounder}            
|F_N(E,\hat{\Theta}) - E_t[F_1(e_i,\Theta^*] |            
\leq            
|F_N(E,\Theta^*) - E_t[F_1(e_i,\Theta^*] | +            
|F_N(E,\hat{\Theta}) - F_N(E,\Theta^*)]|            
\end{equation}            
The first term on the right hand side is the absolute value of the
difference between the sample average of an iid random variable and
its mean value.  This approaches zero almost surely by the strong law
of large numbers and so for sufficiently large $N$ the first term is
less than $\epsilon/2$ with probability greater than $1 - \delta/2$
for any $\epsilon >0$ and $0 < \delta < 1$.  In order to show that the
second term on the right hand side converges to zero in probability,
note that since $F_1(e_i,\Theta)$ is equicontinuous at $\Theta^*$,
given any $\epsilon > 0$ there is a neighbourhood $U$ of $\Theta^*$
within which $|F(e_i,\Theta) - F(e_i,\Theta^*)| < \epsilon/2 $ for any
$e_i$ and $\Theta \in U$.  Therefore, for any set of outcomes $E$ and
$\Theta \in U$, $|F_N(E,\Theta) - F_ N(E,\Theta^*)| = (1/N) |
\sum_{i=1}^N (F_1(e_i,\Theta) - F_1(e_i,\Theta^*)) | \leq(1/N)
\sum_{i=1}^{N} | F_1(e_i,\Theta) - F_1(e_i,\Theta^*)| \leq
\epsilon/2$.  By consistency of the maximum likelihood estimator,
$\hat{\Theta} \in U$ with probability greater than $1 - \delta/2$ for
sufficiently large $N$.  Consequently, $Pr(|F_N(E,\Theta) - F_
N(E,\Theta^*)| > \epsilon/2) < \delta/2$ for sufficiently large $N$.
Putting the bounds on the two terms on the right hand side of
Equation~\ref{eq:bounder} together, and using the union of events
bound we see that for sufficiently large $N$:
\begin{equation}            
\label{eq:result}            
\Pr( |F_N(E,\hat{\Theta}) - E_t[F_1(e_i,\Theta^*] | > \epsilon) <
\delta
\end{equation}            
To complete the proof we can observe that by assumption
$\ln\Pr(e_i|\Theta)$ and its derivatives with respect to $\Theta$ are
equicontinuous at $\Theta^*$ and that $(-1/N) \ln\Pr(E|\Theta)$ and
the various $\tilde{I}_{\mu_1\cdots\mu_i}$ are therefore examples of
the functions of $F$.  Furthermore, $E_t[-\ln\Pr(e_i|\Theta^*)] =
D(t\|\Theta^*) + h(t)$ and
$E_t[\tilde{I}_{\mu_1\cdots\mu_i}|_{\Theta^*}] =
\tilde{J}_{\mu_1\cdots\mu_i}$ under the assumption that derivatives
with respect to $\Theta$ commute with expectations with respect to
$t$.  On applying Equation~\ref{eq:result} to these observations, the
theorem is proved.  $\Box$
       
Note that Lemma~\ref{lemma:conv1} shows that the two leading terms in
the asymptotic expansions of $\chi_E(A) $ and $\chi_N(A)$ approach
each other with high probability.  We will now obtain control over the
subleading terms in these expansions.  Define $c_k$ to be the
coefficient of $1/N^k$ in the asymptotic expansion of $\chi_E(A)$ so
that we can write $\chi_E(A) - N h(t) = N c_{-1} + (d/2) \ln{N} + c_0
+ (1/N) c_1 + (1/N^2) c_2 + \cdots$.  Let $d_k$ be the corresponding
coefficients of $1/N^k$ in the expansion of $\chi_N(A)$.  The $d_k$
are identical to the $c_k$ with each $\tilde{I}$ replaced by
$\tilde{J}$.  We can show that the $c_k$ approach the $d_k$ with high
probability.
\begin{lemma}                
\label{lemma:conv2}                
Let the assumptions made in Lemma~\ref{lemma:conv1} hold and let
$\epsilon > 0 $ and $0 < \delta < 1$.  Then for every intger $k \geq
-1$, there is an $N_k$ such that $\Pr(|c_k - d_k| > \epsilon) <
\delta$.
\end{lemma}                
{\bf Proof: } The coefficient $c_{-1} = (-1/N) \ln\Pr(E|\hat{\Theta})$
has been shown to approach $d_{-1} = D(t\|\Theta^*)$ in probability as
an immediate consequence of Lemma~\ref{lemma:conv1}.  Next we consider
$c_k$ for $k\geq 1$.  Every term in every such $c_k$ can be shown to
be a finite sum over finite products of constants and random variables
of the form $\tilde{I}_{\mu_1\cdots\mu_i}$ and
$\tilde{I}_{\mu\nu}^{-1}$.  We have already seen that
$\tilde{I}_{\mu_1\cdots\mu_i}|_{\hat{\Theta}}\rightarrow
\tilde{J}_{\mu_1\cdots\mu_i}$ in probability.  The
$\tilde{I}_{\mu\nu}^{- 1}|_{\hat{\Theta}}$ are the entries of the
inverse of the empirical Fisher Information
$\tilde{I}_{\mu\nu}|_{\hat{\Theta}}$.  Since the inverse is a
continuous function, and since $\tilde{I}_{\mu\nu} \rightarrow
\tilde{J}_{\mu\nu}$ in probability, $\tilde{I}_{\mu\nu}^{-1}
\rightarrow \tilde{J}_{\mu\nu}^{-1}$ in probability also.  As noted
before, $d_k$ is identical to $c_k$ with each $\tilde{I}$ replaced by
$\tilde{J}$.  Since $c_k$ is finite sum of finite products of random
variables $\tilde{I}$ that converge individually in probability to the
$\tilde{J}$, we can conclude that $c_k \rightarrow d_k$ in
probability.  Finally, we consider $c_0 - d_0=
(-1/2)\ln(\det{J_{ij}(\hat{\Theta})}/\det{\tilde{I}_{\mu\nu}}) -
(-1/2)\ln(\det{J_{ij}(\Theta^*)}/\det{\tilde{J}_{\mu\nu}})$.  We have
shown that $J_{ij}(\hat{\Theta}) \rightarrow J_{ij}(\Theta^*)$ and
$\tilde{I}_{\mu\nu} \rightarrow \tilde{J}_{\mu\nu}$ in probability.
Since the determinant and the logarithm are continuous functions we
conclude that $c_0 - d_0 \rightarrow 0$ in probability. $\Box$
                
We have just shown that each term in the asymptotic expansion of
$\chi_E(A) - h(t)$ approaches the corresponding term in $\chi_N(A)$
with high probability for sufficiently large $N$.  As an easy
corollary of this lemma we obtain the following theorem:
\begin{theorem}                
\label{theorem:conv}            
Let the conditions necessary for lemmas~\ref{lemma:conv1}
and~\ref{lemma:conv2} hold and take $k^\prime \geq k + 1 \geq 0$ to be
integers.  Then let $T_E(A,k,k^\prime)$ consist of the terms in the
asymptotic expansion of $\chi_E(A) $ that are of orders $1/N^k$ to
$1/N^{k^\prime}$.  For example, $T_E(A,4,6) = (1/N^4) c_4 + (1/N^5)
c_5 + (1/N^6) c_6$, using the coefficients $c_k$ defined above.  Let
$T_N(A,k,k^\prime)$ be the corresponding terms in the asymptotic
expansion of $\chi_N(A)$.  Then for any $k$ and $k^\prime$, and for
any $\epsilon >0$ and $0 < \delta < 1$, $\Pr(N^k | T_E(A,k,k^\prime) -
T_N(A,k,k^\prime)| > \epsilon) < \delta$ for sufficiently large $N$.
\end{theorem}                
{\bf Proof: } By definition of $T_E$ and $T_N$, $N^k
|T_E(A,k,k^\prime) - T_N(A,k,k^\prime)| = | \sum_{i=k}^{k^\prime} (c_i
- d_i)/N^{i-k}| \leq \sum_{i=k}^{k^\prime} |c_i - d_i|/N^{i-k}$.  By
Lemma~\ref{lemma:conv2} $|c_i - d_i| \rightarrow 0$ in probability.
Therefore, $N^k |T_E(A,k,k^\prime) - T_N(A,k,k^\prime)|$ is a postive
number that is upper bounded by a finite sum of random variables that
individually converge to zero in probability.  Since the sum is finite
we can conclude that $N^k |T_E(A,k,k^\prime) - T_N(A,k,k^\prime)|$
also converges to zero in probability thereby proving the theorem
$\Box$
                
Note that the multiplication by $N^k$ ensures that the convergence is
not simply due to the fact that every partial sum $T_E(A,k,k^\prime)$
is individually decreasing to zero as the number of outcomes
increases.  Any finite series of terms in the asymptotic expansion of
the logarithm of the Bayesian posterior probability converges in
probability to the corresponding series of terms in the expansion of
the razor.  Theorem~\ref{theorem:conv} precisely characterizes the
sense in which the razor of a model reflects the typical asymptotic
behaviour of the Bayesian posterior probability of a model given the
sample outcomes.
             
      We can also compare the razor to the expected behaviour of
$R_E(A)$ in the true distribution $t$.  Clarke and Barron have
analyzed the expected asymptotics of the logarithm of $t(E)/R_E(A)$
where $t$ is the true distribution, under the assumption that $t$
belongs to the parametric family $A$.(\cite{clarke}) With certain
small modifications of their hypotheses their results can be extended
to the situation studied in this paper where the true density need not
be a member of the family under consideration.  The first modification
is that the expectation values evaluated in Condition 1 of
~\cite{clarke} should be taken in the true distribution $t$ which need
not be a member of the parametric family.  Secondly the
differentiability requirements in Conditions 1 and 2 should be applied
at $\Theta^*$ which minimizes $D(t\|\Theta)$. (Clarke and Barron apply
these requirements at the true parameter value since they assume that
$t$ is in the family.)  Finally, Condition 3 is changed to require
that the posterior distribution of $\Theta$ given $X^n$ concentrates
on a neighbourhood of $\Theta^*$ except for $X^n$ in a set of
probability $o(1/\log N)$.  Under these slightly modified hypotheses
it is easy to rework the analysis of~\cite{clarke} to demonstrate the
following asymptotics for the expected value of $R_E(A)$:
\begin{equation}                              
<- \ln R_E(A)>_{t} - h(t) = N D(t|\Theta^*) + \frac{d}{2}
\ln\left(\frac{N}{2\pi e}\right) -
\frac{1}{2}\ln(\det{J}/\det{\tilde{J}}) - \ln(1/V)
\end{equation}                              
We see that as $N \rightarrow \infty$, $<-\ln R_E(A)>_t -h(t)$ is
equal to the razor up to a constant term $d/2$.  More careful analysis
shows that this term arises from the statistical fluctuations of the
maximum likelihood estimator of $\Theta^*$ around $\Theta^*$.  It is
worth noting that while terms of $O(1)$ and larger in $<\ln R_E(A)>_t$
depend depend at most on the measure (prior distribution) assigned to
the parameter manifold, the terms of $O(1/N)$ depend on the geometry
via the connection coefficients in the covariant derivatives. For that
reason, the $O(1/N)$ terms are the leading probes of the effects that
the geometry of the space of distributions has on statistical
inference in a Baysian setting and so it would be very interesting to
analyze them.  Normally we do not include these terms because we are
interested in asymptotics, but when the amount of data is small, these
correction terms are potentially important in implementing
parsimonious density estimation.  Unfortunately it turns out to be
difficult to obtain sufficiently fine control over the probabilities
of events to extend the expected asymptotics beyond the $O(1)$ terms
and so further analysis will be left to future publications.

\subsection{Relationship to The Minimum Description Length Principle}                                 
In the previous section we have seen that the Bayesian conditional
probability of a model given the data is an estimator of the razor.
In this section we will consider the relationship of the razor to the
Minimum Description Length principle and the stochastic complexity
inference criterion advocated by Rissanen.  The MDL approach to
parameteric inference was pioneered by Akaike who suggested choosing
the model maximizing $\ln\Pr(E|\hat{\Theta}) - d$ with $d$ the
dimension of the model and $\hat{\Theta}$ the maximum likelihood
estimator.(\cite{akaike}) Subsequently, Schwarz studied the
maximization of the Bayesian posterior likelihood for densities in the
Koopman-Darmois family and found that the Bayesian decision procedure
amounted to choosing the density that maximized
$\ln\Pr(E|\hat{\Theta}) - (1/2)d\log{N}$.(\cite{schwartz}) Rissanen
placed this criterion on a solid footing by showing that the model
attaining $\min_{\Theta,d} \{- \log\Pr(E|\Theta) + (1/2)d\log{N}\} $
gives the most efficient coding rate possible of the observed sequence
amongst all universal codes.(\cite{riss84},\cite{riss86}).  In this
paper we have shown that the razor of a model, which reflects the
typical asymptotics of the logarithm of the Bayesian posterior, has a
geometric interpretation as an index of the simplicity and accuracy of
a given model as a description of some true distribution.  In the
previous section we have shown that the logarithm of the Bayesian
posterior can be expanded as:
\begin{eqnarray}                                  
\chi_E(A) = &-\ln{R_E(A)} = -\ln\Pr(E|\hat{\Theta}) + \frac{d}{2}\ln{N}
-\frac{1}{2}\frac{\det{J_{\mu_i\mu_j}(\hat{\Theta})}}{\det{\tilde{I}_{\mu_i\mu_j}}}
\nonumber \\
& -\ln\left[\frac{(2\pi)^{d/2}}{V}\right] + O(1/N) \label{eq:stoch}
\end{eqnarray}                 
with $\hat{\Theta}$ the maximum likelihood parameter and                 
$\tilde{I}_{\mu_1\cdots\mu_i} =   - (1/N)\grad_{\mu_1}\cdots\grad_{\mu_i}                 
\ln{\Pr(E|\Theta)} |_{\hat{\Theta}}$.                 
The term of $O(1/N)$ that we have not explicitly written is the same
as the the corresponding term of the logarithm of the razor
(Equation~\ref{eq:lograzor}) with every $\tilde{J}$ replaced by
$\tilde{I}$.  We recognize the first two terms in this expansion to be
exactly the stochastic complexity advocated by Rissanen as a measure
of the complexity of a string relative to a particular model family.
We have given a geometric meaning to the term $(d/2)\ln N$ in terms of
a measurement of the rate of shrinkage of the volume in parameter
space in which the likelihood of the data is significant.  Given our
results concerning the razor and the typical asymptotics of
$\chi_E(A)$, this strongly suggests that the definition of stochastic
complexity should be extended to include the subleading terms in
Equation~\ref{eq:stoch}.  Indeed, Rissanen has considered such an
extension based on the work of Clarke and Barron and finds that the
terms of $O(1)$ in the expected value of Equation~\ref{eq:stoch}
remove the redundancy in the class of codes that meet the bound on the
expected coding rate represented by the earlier definition of
stochastic complexity.(\cite{rissanen}) Essentially, in coding short
sequences we are less interested in the coding rate and more
interested in the actual code {\it length}.  This suggests that for
small $N$ the $O(1/N)$ terms can be important in determining the ideal
expected codelength but it remains difficult to obtain sufficient
control over the probabilities of rare events to extend the Rissanen's
result to this order.  As mentioned earlier, the metric on the
parameter manifold affects the terms of $O(1/N)$ and therefore these
corrections would be geometric in nature.
                      
Another approach to stochastic complexity and learning that is related to the razor and its                    
estimators has been taken recently by Yamanishi.(\cite{yamanishi})   Let ${\cal H}^d =                    
\{f_\Theta\}$ be a hypothesis class indexed by d-dimensional real vectors $\Theta$.                    
Then,   in a general decision theoretic setting Yamanishi defines the Extended Stochastic                    
Complexity of a model $A$ relative to the data $E$, the class ${\cal H}^d$, and  a loss                    
function $L$ to be:                   
\begin{equation}                   
\label{eq:esc}                   
I(D^N:{\cal H}^d) = -\frac{1}{\lambda}\ln\int d\Theta \: \pi(\Theta) e^{-\lambda                    
\sum_{i=1}^{N} L(D_i:f_\Theta)}                   
\end{equation}                   
where $\lambda > 0$ and $\pi(\Theta)$ is a prior.   Following the work described in this                 
paper he  defines the razor index of $A$ relative to $L$, ${\cal H}^d$  and a given true                    
distribution $p$ to be:                   
\begin{equation}                   
\label{eq:escrazor}                   
I_N(p:{\cal H}^d) = -\frac{1}{\lambda}\ln\int d\Theta \: \pi(\Theta) e^{-\lambda                    
E_p\left[\sum_{i=1}^{N} L(D_i:f_\Theta)\right]}                   
\end{equation}                   
For the case of a loss function $L(E,f_\Theta) = -\ln\Pr(E|\Theta)$,
Equations~\ref{eq:escrazor} and~\ref{eq:esc} reduce to the quantities
$\chi_N(A)$ and $\chi_E(A)$ which are the logarithm of the razor and
its estimator.  Yamanishi shows that if the class of functions ${\cal
H} = \{f_\Theta(X)\}$ has finite Vapnik-Chervonenkis dimension, then
$(1/N) |I(D^N:{\cal H}^d) - I_N(p:{\cal H}^d)| < \epsilon$ with high
probability for sufficiently large $N$.  For the case of a logarithmic
loss function this result applies to the razor and its estimator as
defined in this paper.

\subsection{``Physical'' Interpretation of The Razor}                                 
There is an interesting ``physical'' interpretation of the results
regarding the razor and the asymptotics of Bayes Rule which identifies
the terms in the razor with energies, temperatures and entropies in
the physical sense.  Many techniques for model estimation involve
picking a model that minimizes a loss function $\exp L_\Theta(E)$
where $E$ is the data, $\Theta$ are the parameters and $L$ is some
empirical loss calculated from it.  The typical behaviour of the loss
function is that it grows as the amount of data grows.  In the case of
maximum likelihood model estimation we take $L_\Theta(E) = -N (-
\ln\Pr(E|\Theta))/N$ where we expect $-\ln\Pr(E|\Theta)/N$ to attain a
finite positive limit as $N \rightarrow \infty$ under suitable
conditions on the process generating the data.  In this case we can
make an analogy with physical systems: $N$ is like the inverse
temperature and the limit of $-\ln\Pr(E|\Theta)/N$ is like the energy
of the system.  Maximum likelihood estimation corresponds to
minimization of the energy and in physical terms will be adequate to
find the equilibrium of the system at zero temperature (infinite $N$).
On the other hand we know that at finite temperature (finite $N$) the
physical state of the system is determined by minimizing the free
energy $E - T\,S$ where $T$ is the temperature and $S$ is the entropy.
The entropy counts the volume of configurations that have energy $E$
and accounts for the fluctuations inherent in a finite temperature
system.  We have seen in the earlier sections that terms in the razor
and in the asymptotics of Bayes Rule that account for the simplicity
of a model arise exactly from such factors of volume.  Indeed, the
subleading terms in the extended stochastic complexity advocated above
can be identified with a ``physical" entropy associated with the
statistical fluctuations that prevent us from knowing the ``true"
parameters in estimation problems.
                                 
\subsection{The Natural Parametrization of A Model}                                 
The evaluation of the razor and the relationship to the asymptotics of
Bayes Rule suggest how to pick the ``natural" parametrization of a
model.  In geometric terms, the ``natural" coordinates describing a
surface in the neighbourhood of a given point make the metric locally
flat.  The corresponding statement for the manifolds in question here
is that the natural parametrization of a model in the vicinity of
$\Theta_0$ reduces the Fisher Information $J_{ij}$ at $\Theta_0$ to
the identity matrix.  This choice can also be justified from the point
of view of statistics by noting that for a wide class of parametric
families the maximum likelihood estimator of $\Theta_0$ is
asymptotically distributed as a normal density with covariance matrix
$J_{ij}$.  If $J_{ij}$ is the identity in some parametrization, then
the various components of the maximum likelihood estimator are
independent, identically distributed random variables.  Therefore, the
geometric intuitions for ``naturalness'' are in accord with the
statistical intuitions.  In our context where the true density need
not be a member of the family in question, there is another natural
choice in the vicinity of $\Theta^*$ that minimizes $D(t\|\Theta)$.
We could also pick coordinates in which $\tilde{J}_{ij} =
\grad_{i}\grad_{j}D(t\|\Theta)|_{\Theta^*}$ is reduced to the identity
matrix.  We have carried out an expansion of the Bayesian posterior
probability in terms of $\hat{\Theta}$ which maximizes
$\ln\Pr(E|\Theta)$.  We expect that $\hat{\Theta}$ is asymptotically
distributed as a normal density with covariance $\tilde{J}$.  The
second choice of coordinates will therefore make the components of
$\hat{\Theta}$ independent and identically distributed.

\subsection{Minimum Complexity Density Estimation}                                 
\label{sec:mincomp}                                 
There are numerous close relationships between the work described in
this paper and previous results on minimum complexity density
estimation.  The seminal work of Barron and Cover introduced the
notion of an ``index of resolvability'' which was shown to bound
covergence rates of a very general class of minimum complexity density
estimators.  This class of estimators was constructed by considering
densities which achieve the following minimization:
\begin{equation}                                 
\label{eq:mincomp}                                 
\min_q \left[ L(q) - \log \prod_{i=1}^{N} q(X_i) \right]                                 
\end{equation}                                 
where the $X_i$ are drawn iid from some distribution, $q$ belongs to
some countable list of densities, and the set of $L(q)$ satisfy
Kraft's inequality.(\cite{barron}) Equation~\ref{eq:mincomp} can be
interpreted as minimizing a two stage code for the density $q$ and the
data.  The ``index of resolvability'' $R_n(p)$ of $p$ is constructed
from expectation value in $p$ of Equation~\ref{eq:mincomp} divided by
$N$, the number of samples:
\begin{equation}                                 
\label{eq:index}                                 
R_n(p) = \min_q\left[ \frac{L_N(q)}{N} + D(p\|q) \right]                                 
\end{equation}                                 
where the $L_n$ are description lengths of the densities and $D$ is
the relative entropy. This quantity was shown to bound the rates of
convergence of the minimum complexity estimators.  In a sense the
density achieving the minimization in Equation~\ref{eq:index} is a
theoretical analog of the sample-based minimum complexity estimator
arising from Equation~\ref{eq:mincomp}.
                                 
The work of Barron and Cover starts from the assumption that
description length is the correct measure of complexity in the context
of density estimation and that minimizing this complexity is a good
idea.  They have demonstrated several very general and beautiful
results concerning the consistency of the minimum description length
principle in the context of density estimation.  We also know that
minimum description length principles lead to asymptotically optimal
data compression schemes.  The goal of this paper has been to develop
some alternative intuitions for the practical meaning of simplicity
and complexity in terms of geometry in the space of distributions.
The {\it razor} defined in this paper, like the index of
resolvability, is an idealized theoretical quantity which sample-based
inference schemes will try to approximate.  The razor reflects the
typical order-by-order behaviour of Bayes Rule just as the index of
resolvability reflects the {\it expected} behaviour of the minimum
complexity criterion of Barron and Cover.
                                 
In order to compare the two quantities and their consequences we have
to note that Barron and Cover do not work with families of
distributions, but rather with a collection of densities.
Consequently, in order to carry out inference with a parametric family
they must begin by discretizing the parameter manifold.  The goal of
this paper has been to develop a measure of the simplicity of a family
as a whole and hence we do not carry out such a truncation of a
parameter manifold.  Under the assumption that the true density is
approximated by the parametric family Barron and Cover find that an
optimal discretization of the parameter manifold (see~\cite{wallace}
and~\cite{barron85}) yields a bound on the resolvability of a
parametric model of:
\begin{equation}                                 
\label{eq:resolvbound}                 
R_n(p_\theta) \leq \frac{1}{n}\left( \frac{d}{2} \log{n} +                                 
\log{\frac{\sqrt{J_\theta}}{w(\theta)}} -\frac{d}{2} \log{c_d/e} +                                 
o(1) \right)                                  
\end{equation}                                 
where $\theta$ is the true parameter value, $J(\theta)$ is the Fisher
Information at $\theta$, $w(\theta)$ is a prior density on the
parameter manifold and $c_d$ arises from sphere-packing problems and
is close to $2\pi e$ for large $d$.  The asymptotic minimax value of
the bound is attained by choosing the prior $w(\Theta)$ to be Jeffreys
prior $\sqrt{\det{J}}/\int d\mu(\Theta) \sqrt{\det{J}}$.\footnote{This
is one of the coding theoretic justifications for the choice of
Jeffreys prior.  We have provided a novel discussion of the choice of
that pior in this paper.}  We see that aside from the factor of $1/N$,
the leading terms reproduce the logarithm of the razor for the case
when the true density is infinitesimally distant from the parameter
manifold in relative entropy sense so that $D(t\|\theta^*) = 0$.
Barron and Cover use this bound to evaluate convergence rates of
minimum complexity estimators.  In contrast, this paper has presented
the leading terms in an asymptotically exact expansion of the razor as
an abstract measure of the complexity of a parametric model relative
to a true distribution.  I begin by studying what the meaning of
``simplicity'' should be in the context of Bayes Rule and the geometry
of the space of distributions, and arrive at results that are closely
related to the minimum complexity scheme. Saying the models with
larger razors are preferred is asymptotically equivalent to saying the
models with a lower resolvability (given an optimal discretization)
are preferred.  However, we see from a comparison of the logarithm of
the razor and Equation~\ref{eq:resolvbound}, that the resolvability
bound is a truncation of the series expansion of the log razor which
therefore gives a finer classification of model families.  The
geometric formulation of this paper leads to interpretations of the
various terms in the razor that give an alternative understanding of
the terms in the index of resolvability that govern the rate of
convergence of minimum complexity estimators.  We have also given a
systematic scheme for evaluating the razor to all orders in $1/N$.
This suggests that the results on optimal discretizations of parameter
manifolds used in the index of resolvability should be extended to
include such sub-leading terms.(\cite{wallace},\cite{barron85})

\section{Conclusion}                                 
\label{sec:conclusion}                                 
In this paper we have set out to develop a measure of complexity of a
parametric distribution as a description of a particular true
distribution.  We avoided appealing to the minimum description length
principle or to results in coding theory in order to arrive at a more
geometric understanding in terms of the embedding of the parametric
model in the space of probability distributions.  We constructed an
index of complexity called the razor of a model whose asymptotic
expansion was shown to reflect the accuracy and the simplicity of the
model as a description of a given true distribution.  The terms in the
asymptotic expansion were given geometrical interpretations in terms
of distances and volumes in the space of distributions.  These
distances and volumes were computed in a metric and measure given by
the Fisher Information on the model manifold and the square root of
its determinant.  This metric and measure were justified from a
statistical and geometrical point of view by demonstrating that in a
certain sense a uniform prior in the space of distributions would
induce a Fisher Information (or Jeffreys) prior on a parameter
manifold.  More exactly, we assumed that indistinguishable
distributions should not be counted separately in an integral over the
model manifold and that there is a ``translation invariance'' in the
space of distributions.  We then showed that a Jeffreys prior can be
rigorously constructed as the continuum limit of a sequence of
discrete priors consistent with these assumptions.  A technique of
integration common in statistical physics was introduced to facilitate
the asymptotic analysis of the razor and it was also used to analyze
the asymptotics of the logarithm of the Bayesian posterior.  We have
found that the razor defined in this paper reflects the typical
order-by-order asymptotics of the Bayesian posterior probability just
as the index of resolvability of Barron and Cover reflects the
expected asymptotics of the minimum complexity criterion studied by
those authors.  In particular, any finite series of terms in the
asymptotic expansion of the logarithm of the Bayesian posterior
converges in probability to the corresponding series of terms in the
asymptotic expansion of the razor.  Examination of the logarithm of
the Bayesian posterior and its relationship to the razor also
suggested certain subleading geometrical corrections to the expected
asymptotics of Bayes Rule and corresponding corrections to stochastic
complexity defined by Rissanen.
                 
\section{Acknowledgments}
I would like to thank Kenji Yamanishi for several fruitful
conversations.  I have also had useful discussions and correspondence
with Steve Omohundro, Erik Ordentlich, Don Kimber, Phil Chou and Erhan
Cinlar.  Finally, I am grateful to Curt Callan for his support for
this investigation and to Phil Anderson for helping with travel funds
to the 1995 Workshop on Maximum Entropy and Bayesian Methods.  This
work was supported in part by DOE grant DE-FG02-91ER40671.

\end{document}